\documentclass[aps,showpacs,onecolumn,preprint,amsmath,amssymb,superscriptaddress,intlimits,longbibliography,showkeys]{revtex4-1}

\usepackage{graphicx}
\usepackage{subfigure}
\usepackage{xcolor}
\usepackage{epstopdf}
\usepackage{amssymb}
\usepackage{amsmath}
\usepackage{amsfonts}
\usepackage{dcolumn}
\usepackage{bm}
\usepackage{soul}
\usepackage{array}
\usepackage{physics}

\newcommand{\vk}{{\bf k}}
\newcommand{\vQ}{{\bf Q}}
\newcommand{\vq}{{\bf q}}

\newcolumntype{C}[1]{>{\centering\let\newline\\\arraybackslash\hspace{0pt}}m{#1}}
\newcommand{\Lmin}{\Lambda_\mathrm{min}}

\providecommand{\keywords}[1]
{
  \small	
  \textbf{\textit{Keywords---}} #1
}

\begin{document}

\title{Exciton-phonon coupling induces new pathway for ultrafast intralayer-to-interlayer exciton transition and interlayer charge transfer in WS$_2$-MoS$_2$ heterostructure: a first-principles study}

\author{Yang-hao Chan}
\email{yanghao@gate.sinica.edu.tw}
\affiliation{Institute of Atomic and Molecular Sciences, Academia Sinica, Taipei 10617, Taiwan}
\affiliation{Physic Division, National Center of Theoretical Physics, Taipei 10617, Taiwan}

\author{Mit H. Naik}
\affiliation{Department of Physics, University of California, Berkeley, CA, 94720-7300, USA}
\affiliation{Materials Science Division, Lawrence Berkeley National Laboratory, Berkeley, CA, 94720, USA}

\author{Jonah B. Haber}
\affiliation{Department of Physics, University of California, Berkeley, CA, 94720-7300, USA}
\affiliation{Materials Science Division, Lawrence Berkeley National Laboratory, Berkeley, CA, 94720, USA}

\author{Jeffrey B. Neaton}
\affiliation{Department of Physics, University of California, Berkeley, CA, 94720-7300, USA}
\affiliation{Materials Science Division, Lawrence Berkeley National Laboratory, Berkeley, CA, 94720, USA}

\author{Steven G. Louie}
\affiliation{Department of Physics, University of California, Berkeley, CA, 94720-7300, USA}
\affiliation{Materials Science Division, Lawrence Berkeley National Laboratory, Berkeley, CA, 94720, USA}

\author{Diana Y. Qiu}
\email{diana.qiu@yale.edu}
\affiliation{Department of Mechanical Engineering and Materials Science, Yale University, New Haven, CT 06520}

\author{Felipe H. da Jornada}
\email{jornada@stanford.edu}
\affiliation{Department of Materials Science and Engineering, Stanford University, Stanford, CA 94305, USA}

\begin{abstract}
Despite the weak, van-der-Waals interlayer coupling, photoinduced charge transfer vertically across atomically thin interfaces can occur within surprisingly fast, sub-50fs timescales. Early theoretical understanding of the charge transfer is based on a noninteracting picture, neglecting excitonic effects that dominate the optical properties of such materials. Here, we employ an \textit{ab initio} many-body perturbation theory approach which explicitly accounts for the excitons and phonons in the heterostructure. Our large-scale first-principles calculations directly probe the role of exciton-phonon coupling in the charge dynamics of the WS$_2$/MoS$_2$ heterobilayer. We find that the exciton-phonon interaction induced relaxation time of photo-excited excitons at the $K$ valley of MoS$_2$ and WS$_2$ is 67 fs and 15 fs at 300 K, respectively, which sets a lower bound to the intralayer-to-interlayer exciton transfer time and is consistent with experiment reports. We further show that electron-hole correlations facilitate novel transfer pathways which are otherwise inaccessible to non-interacting electrons and holes. 
\end{abstract}


\date{\today}
\maketitle
The freedom to stack quasi-two-dimensional (quasi-2D) van der Waals (vdW) materials, with little restrictions imposed by lattice matching, introduces a vast parameter space for designing and engineering device properties, as well as exploring new physics by tuning electron correlations and order parameters through proximity effects~\cite{Geim2013}. In particular, stacked layers of transition metal dichalcogenides (TMDs) have attracted a lot of interest owing to the unique interplay of spin, valley, and optical chirality properties and promising applications in optoelectronic devices~\cite{Rivera2018}. TMD heterostructures can form semiconductors with type II band-alignment such that the valence and conduction band edge at their respective $K$ points in momentum space have wavefunction characters coming from distinct layers. In this setup, optically excited excitons, which are typically formed by electrons and holes in the same layer, can scatter to lower-energy stable interlayer excitons with longer recombination times. The time scale and microscopic mechanism behind such exciton transfer processes are of fundamental interest and crucial for applications ranging from energy harvesting applications to quantum information sciences~\cite{Kim2017}.

Recent pump-probe experiments have suggested that for the WS$_2$/MoS$_2$ heterobilayer, such inter-layer charge transfer can take place on a time scale of less than 50 fs~\cite{Hong2014}. This exceptionally short charge transfer time is surprising, since it is well-understood that the valence band maximum (VBM) and conduction band minimum (CBM) from individual layers -- at the $K$ point -- are only weakly hybridizing. Moreover, phonon modes of the heterostructure also display negligible coupling, apart from very long wavelength acoustic modes. Further, there is a momentum mismatch between the intralayer and interlayer exciton that would prevent a direct Coulomb-mediated charge transfer between the lowest exciton states of the two kind. Later experiments on this system show that this ultrafast transfer is stacking angle independent~\cite{Yu2015,Ji2017,Zhu2017,Jin2018} and has a weak 
dielectric-environment~\cite{Zhou2019} dependence. Despite the intense experimental and theoretical efforts~\cite{Rigosi2015,WangYong2017}, it has so far remained difficult to disentangle the complex experimental observations in vdW heterostructures~\cite{Schaibley2016,Chen2016,Nagler2017,Bian2020,Policht2021} in terms of the competing interaction mechanisms owing in part to the lack of quantitatively predictive \textit{ab initio} theories that treat vibrational and electronic correlation effects on the same footing.

Earlier theoretical work on WS$_2$/MoS$_2$ heterostructure based on time-dependent density functional theory (TD-DFT) suggested that charge transfer is initiated by coherent charge oscillations and completed by electron-phonon interactions~\cite{Wang2016}. Non-adiabatic molecular dynamics (NAMD) approaches provided evidence supporting the role of quantum coherence \cite{Long2016} and electron-phonon interactions~\cite{Zheng2017,Zheng2017nano,ZhangMeng2017,Wang2021,Zeng2021}. Excitonic effects have also been incorporated in recent NAMD studies by combining the sampling of the atomic motion from Born-Oppenheimer molecular dynamics with electronic excited-state calculations. The different excited-state potential energy surfaces are approximated either within TD-DFT, using parametrized range-separated hybrid functionals~\cite{Liu2020}, or within many-body perturbation theory by solving the Bethe-Salpeter equation (BSE)~\cite{Jian2021}. In a recent work, a new excitonic channel for intralayer-to-interlayer charge transfers was discovered with TD-aGW calculations~\cite{Chen2023}. Overall, these calculations uncovered important aspects on the microscopic mechanism beyond independent-particle interlayer charge transfer, and suggest that a two-step relaxation process occurs in the vdW heterostructure, with excitonic effects aiding in the process~\cite{Liu2020}.

However, while promising, the computational demand of most aforementioned methods is quite steep, and one typically sacrifices the description of electronic correlations to obtain the coupled description of electrons and phonons. For instance, despite earlier pioneering efforts, NAMD calculations are often still restricted to small supercells, which makes the spectrum of excitonic states artificially sparse~\cite{Qiu2016}, in particular around the energy of intralayer excitons, and may lead to qualitatively different exciton decay pathways.

In this letter, we study exciton transfer (i.e., the rate of scattering from one exciton state to other exciton states via phonons) in a WS$_2$/MoS$_2$ TMD bilayer heterostructure including both electron-hole and exciton-phonon~\cite{Antonius2017,Chen2020,huang2021} interactions fully from first principles within the framework of many-body perturbation theory. Our computed relaxation time of the two lowest-energy intralayer excitons (the A excitons in the two layers) shows that exciton-phonon couplings are capable of inducing ultrafast charge transfer. Moreover, in contrast to the two-step transfer pathway proposed in earlier NAMD studies, we find a direct charge transfer pathway enabled by electron-hole correlations and intravalley scattering, with scattering rates in agreement with the sub-50~fs bleaching of optical signatures seen in experiments~\cite{Hong2014,Chen2016,Ji2017,Jin2018}. Our direct \textbf{k}-space-resolved exciton and electron dynamics analysis offers a detailed picture of the charge-transfer process.

\begin{figure}[t]
     \centering
     \includegraphics[width=.9\textwidth]{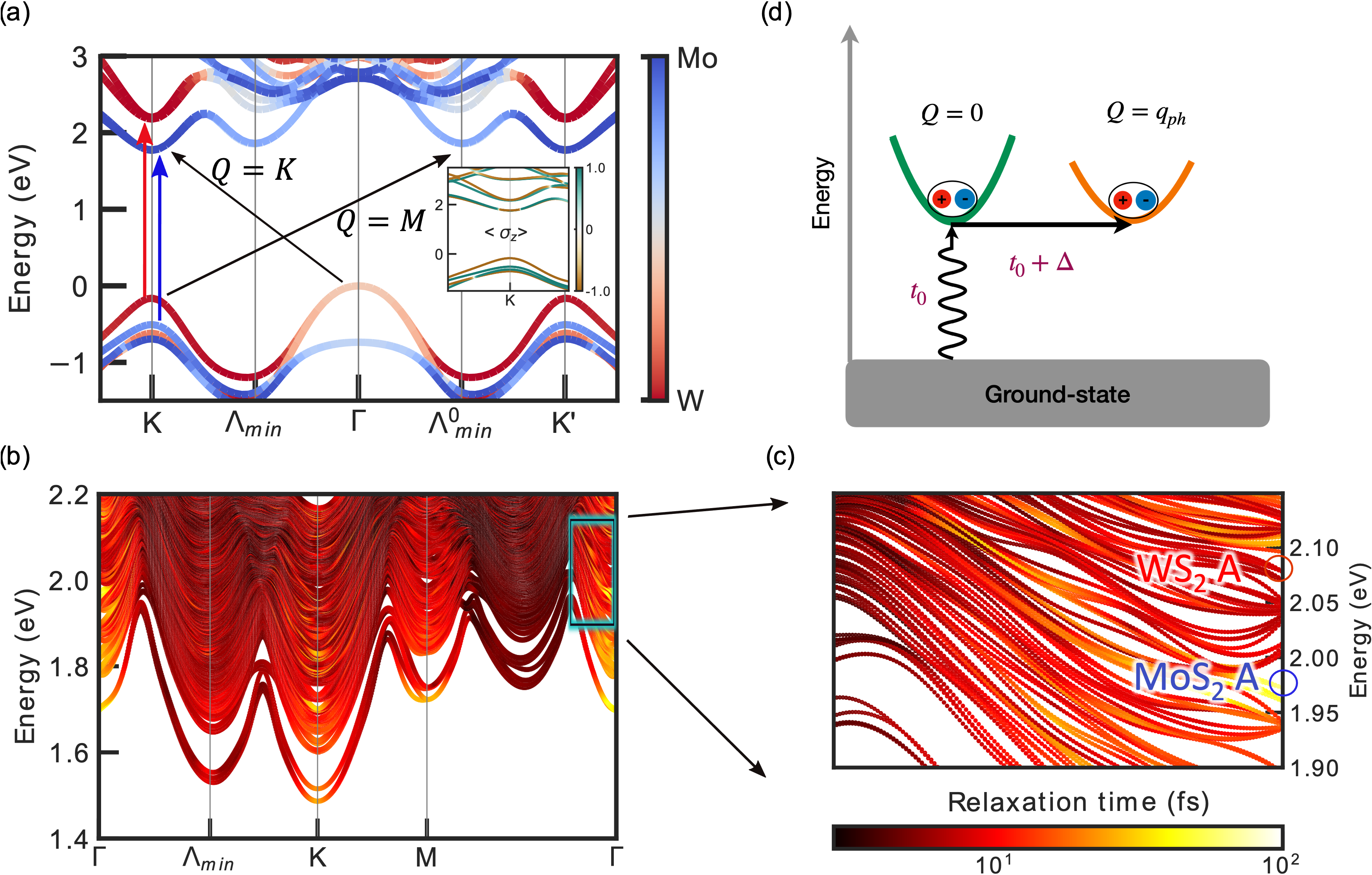}
\caption{(a) Electron band dispersion of the WS$_2$/MoS$_2$ heterostructure overlaid with a color scale proportional to the contribution of Mo and W atoms to the projected component of the wavefunction squared. Interband transitions corresponding to excitons with center of mass momentum $\mathbf{Q}=K$, and $\mathbf{Q}=M$ are shown with the labeled arrows. The blue and red arrows indicate transition corresponding to MoS$_2$ and WS$_2$ A excitons with $\mathbf{Q}=\Gamma$, respectively. The inset shows a colormap of z-component of the spin expectation values of states near the $K$ valley. (b) Exciton dispersion of both bound and resonant states of the WS$_2$/MoS$_2$ heterobilayer along a path of exciton center of mass momentum. Color indicates the value of the relaxation time due to exciton-phonon coupling at 300 K. (c) A zoom-in of panel (b) around the MoS$_2$ and WS$_2$ A excitons, which are indicated by blue and red circles, respectively. (d) Schematic representation of phonon-mediated exciton scattering process. A $\vQ=\Gamma$ exciton forms at $t_0$ and relaxes to an exciton with $\vQ=\vq_{ph}$ at a later time $t_0+\Delta$ due to exciton-phonon interactions.}
\label{fig:Relax}
\end{figure}

We focus on the energetically favorable H$^M_h$ stacking WS$_2$/MoS$_2$ heterostructure, with a twist angle of $60^\circ$. Fig.~\ref{fig:Relax} (a) shows the electronic band structure calculated with the G$_0$W$_0$ approach as implemented in the BerkeleyGW software package~\cite{Hybertsen1986,Deslippe2012}.
The band structure at the $K$ valley clearly exhibits a typical type II band alignment. The first (topmost) valence band at the $K$ valley is of WS$_2$ character while the second valence band is of MoS$_2$ character. The next two valence bands also follow the same order. The lowest two conduction bands at the $K$ valley both have MoS$_2$ character but opposite spin. The next two conduction bands are of WS$_2$ character. The system is an indirect band-gap semiconductor with the valence band maximum (VBM) at $\Gamma$ (which is of hybridized characters of the two layers) and the conduction band minimum (CBM) at $K$. The direct band gap is 1.93 eV at the $K$ and $K'$ valleys, where the valence band is about 160 meV lower than the VBM. This energy landscape provides a relaxation path for holes generated in the $K$ valley. Furthermore, the mix of Mo and W orbital character at the $\Gamma$ and $\Lambda_{min}$ valleys suggest an intermediate state that can mediate an intralayer to interlayer charge transfer pathway. In panel (b) we show the exciton dispersion by solving the Bethe-Salpeter equation (BSE) within the GW-BSE method~\cite{Hybertsen1986,Rohlfing2000} for excitons with finite center-of-mass (COM) momentum $\vQ$~\cite{Qiu2015}. Due to the indirect band gap nature, the lowest-energy exciton has a finite COM momentum of $\vQ=K$ and is of interlayer character. Our calculations show that the first intralayer bright exciton in the MoS$_2$ layer has an energy of 1.97 eV and the first intralayer WS$_2$ bright exciton, which must have $\vQ=0$, is located at 2.09 eV. We will refer to these bright excitons as the MoS$_2$ and WS$_2$ A exciton, respectively. These peaks agree well with previous experiment \cite{Hong2014}, within 100~meV, and calculations~\cite{Torun2018}. We note that both bright excitons lie above the quasiparticle continuum at 1.93 eV of the lower-energy interlayer excitons. Although bound interlayer excitons are abundant below the continuum they only couple weakly to light. In the inset of Fig.~\ref{fig:Relax} (a), we show the quasiparticle bandstructure overlaid with the spin expectation value. The spin-split bands at the $K$ valley are a consequence of strong spin-orbit coupling. Due to the spin-polarization near the $K$ valley, the bright MoS$_2$ A exciton consists of electrons from the first conduction band and holes from the second valence band, counted from the Fermi energy downward, and the WS$_2$ A exciton consists of electrons from the fourth conduction band and holes from the first valence band. We emphasize that the internal spin structures of each exciton is also relevant to selection rules for exciton-phonon couplings, which is crucial to understand exciton scattering pathways~\cite{Chan2023}.

Exciton-phonon coupling matrix elements can be written as a contraction of electron-phonon coupling matrix elements with the exciton envelope function given in previous works~\cite{Toyozawa1958,Antonius2017,Chen2020,Chan2023}. 
The lowest-order self-energy due to exciton-phonon couplings reads 
\begin{align}
\Sigma^{S\vQ}(\omega) = \frac{1}{\mathcal{N}_\mathbf{q}}\sum_{S',\vq,\nu,\pm}\frac{|G_{S'S\nu}(\vQ,\vq)|^2(N_{\nu\vq} +\frac{1}{2}\pm\frac{1}{2})}{\omega-E^{S'\vQ+\vq}\mp\omega_{\nu\vq}+i\eta},
\label{Eq:sediag}
\end{align}
where $E^{S'\vQ+\vq}$ is the exciton energy of a state with quantum number $S'$  and a COM momentum $\vQ+\vq$; $\omega_{\nu\vq}$ is the phonon frequency; $N_{\nu \vq}$ is the Bose-Einstein occupation factor associated with the phonon, $ \mathcal{N}_{\mathbf{q}}$ is the number of the wavevectors sampled in the Brillouin zone (BZ); and finally, $G_{S'S\nu}(\vQ,\vq)$ is the exciton-phonon coupling matrix element encoding the probability amplitude for an exciton initially in state ($S, \vQ$) to scatter to state ($S', \vQ+\vq$) through the emission or absorption of a phonon $(\nu, \vq)$. In Fig.~1 (d) we illustrate a typical scattering process which brings an exciton with zero COM momentum to a COM momentum of $\vq$.

\begin{figure*}[t]
     \centering
     \includegraphics[width=\textwidth]{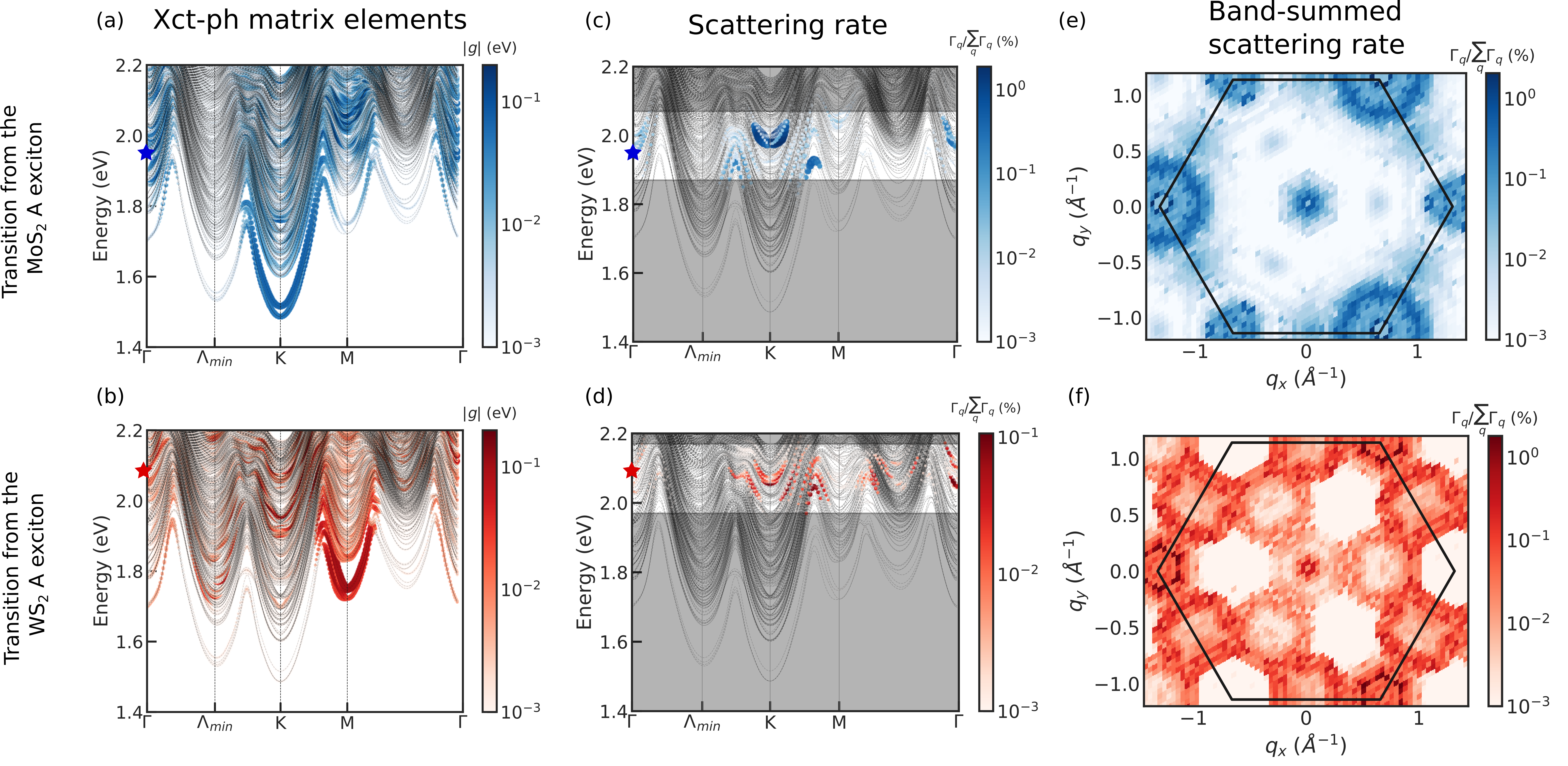}
\caption{Exciton-phonon coupling strength (panel (a) and (b)) and phonon momentum \vq-resolved contribution to the total scattering rate for the MoS$_2$ A exciton (blue star and panel (c),(e)) and WS$_2$ A exciton (red star and panel (d),(f)). (a) and (b) are the color maps of band-resolved amplitude of the exciton-phonon coupling matrix element along a high symmetry path. The color scale and symbol sizes in (c) and (d) indicate the normalized contribution of the scattering rate from the starred state to other states summed over phonon modes. (e) and (f) show the normalized contribution to the scattering rate of the starred state from different phonon momenta \vq. Shaded region in (c) and (d) indicate the energy range where scatterings from A excitons are forbidden due to energy conservation.}
\label{fig:analysis}
\end{figure*}

Exciton scattering rates (inverse of the relaxation time) are evaluated from the the imaginary part of on-shell exciton self-energy in Eq.~\ref{Eq:sediag}. 
We emphasize this quantity gives the rate of an exciton being scattered from one state to all other states via the exciton-phonon coupling, and it is not between two specific states.
In Fig.~1 (b) we show a color map of the exciton-phonon relaxation times in the heterostructure at 300 K. Our calculation shows that the MoS$_2$ A exciton has a relaxation time of 67 fs while WS$_2$ A exciton has a shorter relaxation time of 15 fs. If we take the computed relaxation time to be dominantly due to transitions to exciton states of interlayer character (see below), our results are consistent with the interpretation of the ultrafast charge transfer time observed in the experiments \cite{Hong2014,Chen2016,Ji2017,Jin2018} and with that exciton-phonon interactions dictate the observed ultrafast optical response. Fig.~1 (d) shows a close-up of the region where the two A excitons are located. Since both excitons are in the continuum of the interlayer excitations, one would expect that the available phase space for exciton-phonon scattering is abundant and their relaxation times would be comparable. Yet, we find that the MoS$_2$ A exciton has a relatively long lifetime compared to WS$_2$ A exciton. To understand this result, we analyze momentum and state-resolved exciton-phonon couplings in Fig.~\ref{fig:analysis}. In Fig.~\ref{fig:analysis} (a) and (b) we show color maps of the absolute value of band resolved exciton-phonon coupling strength between the MoS$_2$ A exciton and WS$_2$ A exciton and other states summed over all phonon branches. We observe that both excitons are strongly coupled to $\vQ=M$ excitons. Excitons with a COM momentum near $\vQ=K$ also have appreciable coupling matrix elements. $\vQ=M$ excitons consist of electrons at $\Lmin'$ and holes at $K$, which we will denote as $(c\Lmin',vK)$ pairs while $\vQ=K$ excitons can be either $(cK',vK)$ or $(cK,v\Gamma)$ excitons.

The analysis based purely on exciton-phonon coupling matrix elements does not present a full picture of the relaxation time, which also includes energy conservation (imaginary part of Eq.~\ref{Eq:sediag}).
In Fig.~\ref{fig:analysis} (c) and (d) we show exction states resolved contributions to the scattering rate of the MoS$_2$ (blue star) and WS$_2$ (red star) A excitons along a high symmetry path, respectively. Due to energy conservation conditions, only excitons in the energy window within one phonon frequency can contribute. Hence, for the both layers, the A exciton relaxation time is dominated by scattering from $\vQ=\Gamma$ to $\vQ=K$ rather than $\vQ=\Gamma$ to $\vQ=M$, which corresponds to the electron state of the exciton scattering between the $K$ and $K'$ valleys.

We further analyze the scattering rate contributions resolved in the full BZ, as shown in Fig.~\ref{fig:analysis} (e) and (f), which reveals important scattering channels for the MoS$_2$ and WS$_2$ A excitons, respectively, to any excitonic states via absorption and emission of phonons with wavevector $\vq$. These scattering rates in panels (e) and (f) correspond to the same initial states highlighted with stars in Fig.~\ref{fig:analysis} (c) and (d), respectively. We first study the scattering pathway for an initially excited A exciton in MoS$_2$. We observe that the pattern of contributions to the scattering rate shows a double-ring structure around the $K$ valley. For the outer rings, our analysis shows that the final excitons have electrons and holes located mostly at the $K$ and $\Gamma$ points, respectively, (See \cite{SM}) indicating that the exciton-phonon scattering was mostly due to a change in the hole momenta. Because the valence states near $\Gamma$ are layer-hybridized, we conclude that this scattering pathway is important for the ultrafast charge transfer observed in TMDC heterobilayers.

On the other hand, scattering events associated with phonon wavevectors $\vq$ in the inner ring around $K$, are mostly associated with excitons wherein the electrons and holes are distributed at the $K$ and $K'$ valleys, respectively (See \cite{SM}).  Importantly, we obverse that the final holes change from their original character -- mostly MoS$_2$-like at the second valence band at $K$ --, to mostly WS$_2$-like at the VBM at $K'$. 
This channel indicates that the MoS$_2$ A exciton can directly couple to interlayer excitons and cause Pauli blocking of the states associated with the WS$_2$ A peak, causing photobleaching of the WS$_2$ absorption signal after the optical pumping of the MoS$_2$ A exciton.

We find that the one-step scattering-event channel associated with the inner ring gives a bleaching time of the WS$_2$ exciton of about 250~fs. On the other hand, we also study a previously discussed two-step scattering mechanism, by which MoS$_2$ A excitons are first scattered to excitons wherein the holes are distributed around to the VBM at $\Gamma$ (the outer ring process), which subsequently scatters to interlayer excitons. We find that such a two-step process gives a bleaching time of the WS$_2$ exciton of about 200~fs. Similar scattering time was reported in Ref.~\cite{Wang2021} for MoSe$_2$/WSe$_2$ bilayer.  In Ref~\cite{Zheng2017}, using non-adiabtic molecular dynamics simulations but neglecting excitonic effects, authors conclude that holes placed in the second valence band at $K$, with initial MoS$_2$ character, relax via both pathways detailed above on timescales of a few hundred femptoseconds. These individual time scales compare well, though are a bit longer than the bleaching time of the WS$_2$ A exciton observed in experiment~\cite{Hong2014}, and, altogether, strengthen the case that multiple scattering mechanisms are important for ultrafast charge transfer in bilayer MoS$_2$/WS$_2$. We further give a detailed discussion on the temperature dependence of this effect in the supporting information~\cite{SM}.

Next, we focus on the dynamics of the WS$_2$ A exciton. Fig.~\ref{fig:analysis} (f) shows that the WS$_2$ A exciton can scatter by emission or absorption of phonon wavevectors $\vq$ over a larger region of the BZ. In particular, scatterings via phonons with $\vq\sim K$, $\vq\sim M$, $\vq\sim 0$ and $\vq\sim \Lmin'$ are all viable. Overall, exciton-phonon coupling matrix elements of the MoS$_2$ and WS$_2$ A excitons are of the same order of magnitude; the larger scattering phase space of WS$_2$ A exciton results in its shorter relaxation time. 

After showing that exciton-phonon scattering can be fast enough to explain ultrafast charge transfer observed in a WS$_2$/MoS$_2$ heterostructure, an important follow-up question is to microscopically understand how excitons relax after being excited by an optical field and what is the role of electron-hole correlation in that relaxation. Since a full pump-probe simulation including phonons and the exciton relaxation dynamics remains challenging from \textit{ab initio} many-body perturbation theory, we compute the population redistribution rate of each independent-particle orbital of the many-electron system as an indication of how the carrier population changes as the excitons relax. We define the band- and $\mathbf{k}$-resolved quasiparticle redistribution rate of a quasi-electron state $c\mathbf{k}$ with an initially occupied $\mathrm{A}$ exciton as 
\begin{align}
\tau_{c\mathbf{k}}^{-1}(\mathrm{A})=&\frac{d\left\langle \phi\left|c_{c\mathbf{k}}^{\dagger}c_{c\mathbf{k}}\right|\phi\right\rangle}{dt}\nonumber\\
=&2\pi\sum_{v\mathbf{q}\nu}\sum_{E^S=E^{S'}}\delta\left(E^{S\mathbf{q}}-E^{\mathrm{A}}+\hbar\omega_{\mathbf{q}\nu}\right)G_{S'\mathrm{A}\nu}^{*}(0,\mathbf{q})G_{S\mathrm{A}\nu}(0,\mathbf{q})A_{cv\mathbf{k}}^{S'\mathbf{q},*}A_{cv\mathbf{k}}^{S\mathbf{q}},
\label{Eq:rate}
\end{align}
where the expectation values are taken over an evolved state $\left|\phi\right\rangle$ starting from either the MoS$_2$ or WS$_2$ A exciton. $A^{S\mathbf{q}}$ is the exciton envelope function of a state ($S$,$\mathbf{q}$). Here, the scattered exciton COM momentum coincides with phonon momentum $\mathbf{q}$ since initial A excitons have zero COM momentum. A similar expression of the redistribution rate for valence electrons, along with its derivation, is given in the supporting information \cite{SM}. In the case with non-degenerate exciton bands at specific $\vq$, Eq.~\ref{Eq:rate} reduces to
\[
\tau_{c\mathbf{k}}^{-1}(\mathrm{A})
=2\pi\sum_{v\mathbf{q}S\nu}\delta\left(E^{S\mathbf{q}}-E^{\mathrm{A}}+\hbar\omega_{\mathbf{q}\nu}\right)\left|G_{S\mathrm{A}\nu}(0,\mathbf{q})\right|^2 \left|A_{cv\mathbf{k}}^{S\mathbf{q}}\right|^2,
\]
 which can be understood as the exciton-phonon scattering rate weighted by the electron or hole components of the exciton envelope functions. In contrast to the independent-particle picture of electron-phonon scattering, we can see from the above expression that different scattering pathways become possible due to the nonlocal distribution of electron or hole amplitude in reciprocal space for excitonic states.

To see the effects of electron-hole interactions in the electron relaxation dynamics, we compare the computed quasi-particle redistribution rate with and without electron-hole interactions in Fig. 3. We show the $\vk$- and band-resolved distribution of electrons and holes in the bilayer due to the presence of an initial photoexcited A exciton on MoS$_2$ and WS$_2$ in panels (a) and (e), respectively. We show the phonon-induced evolution in occupation of different quasiparticle states, including electron-hole interactions, by plotting $\tau_{\alpha\mathbf{k}}^{-1}(\mathrm{A})$ in panels (c) and (g), respectively. From Fig.~3 (c), it is clear that, for an initial excitation of the MoS$_2$ A exciton, electrons mostly scatter within the same valley. Intervalley scattering to $K'$ and a remote $\Lmin'$ occurs with less probability. On the other hand, holes -- initially at the second-highest valence band -- scatter to both (i) the VBM at $\Gamma$ and (ii) the two highest-bands in the $K'$ valley. Since the $\Gamma$ valley has a mixed character of both W and Mo atoms and the first band in the $K'$ valley is of W character, both scattering processes (i) and (ii) result in interlayer charge transfers.

We also perform the corresponding calculations without excitonic effects (i.e., considering only electron-phonon interactions). If we start from the lowest-energy vertical transition (intralayer interband transitions) on MoS$_2$ -- blue stars in panel (b) -- the corresponding quasiparticle scattering rate due to electron-phonon interactions are shown in panel (d). The scattering rate is much more limited in this case owing to the stricter energy-momentum conservation conditions in the non-interacting case. In contrast, excitons with energy close to the MoS$_2$ A exciton can have wide variety of energy and momentum distributions given their different possible internal structure, which allows for the coupling of a variety of states in the BZ. In particular, while holes transfer directly to the VBM at $\Gamma$ and the $K'$ valley, it takes a secondary scattering event for this interlayer charge transfer if  electron-hole interactions are not taken into account.

\begin{figure}
    \centering
    \includegraphics[width=\textwidth]{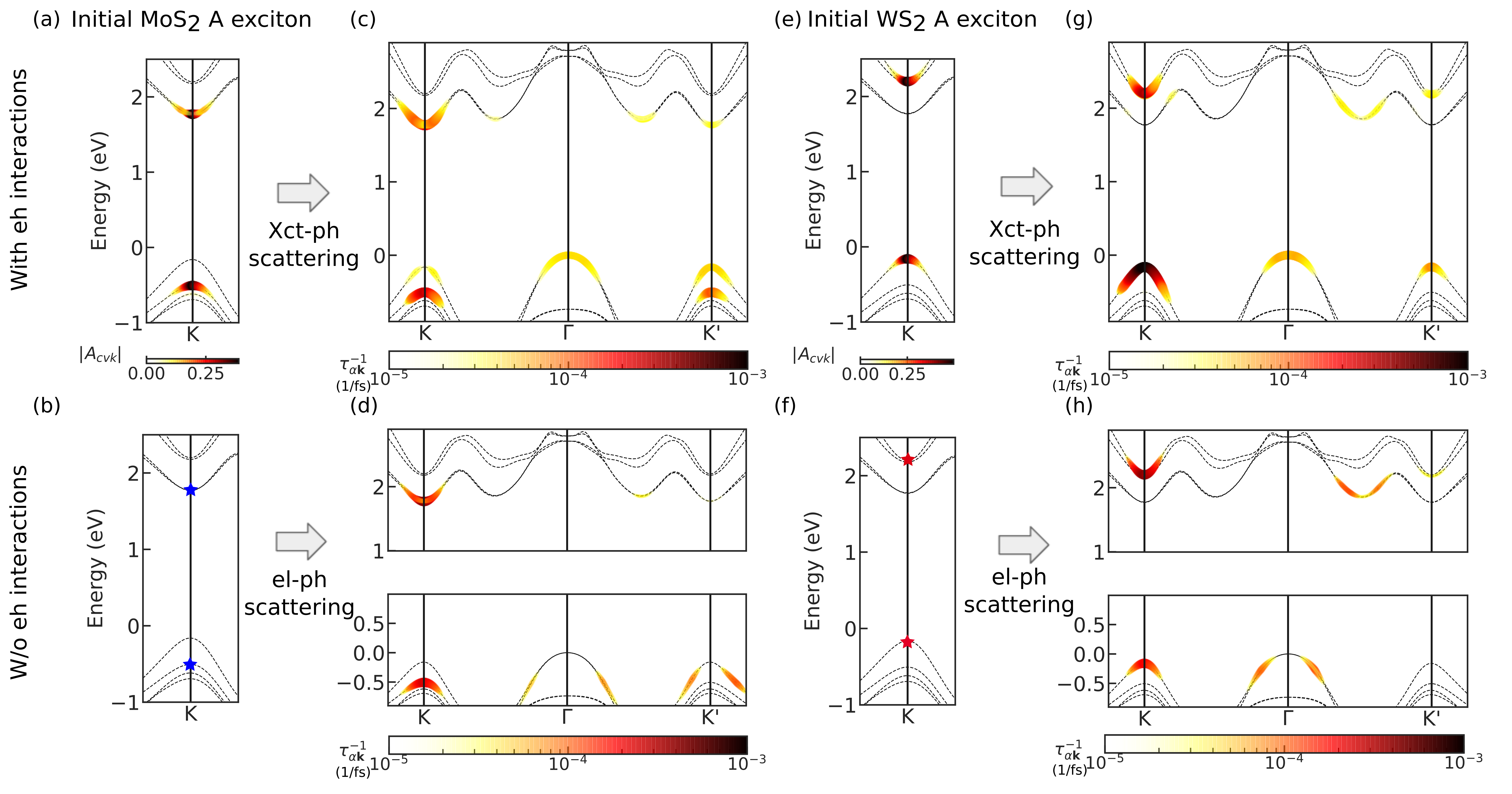}
    \caption{Evolution of the electrons and holes in bilayer MoS$_2$/WS$_2$ due to exciton-phonon interactions. Initial distribution of electrons and holes in the intralayer A exciton from MoS$_2$ (a) and WS$_2$ (e). Band- and k-resolved quasiparticle redistribution rate, $\tau_{\alpha\mathbf{k}}^{-1}(\mathrm{A})$, due to exciton-phonon interactions, after the initial excitation of MoS$_2$ (c) or WS$_2$ (g). Blue (red) stars in panel (b) ((f)) represent the initial position of a free electron and hole at the $K$ valley. (d) and (h) show the inverse of the redistribution time of independent electrons and holes due to electron-phonon scattering starting from the initial distribution in (b) and (f), respectively.}
    \label{fig:MoS2-rate}
\end{figure}

For the initially excited WS$_2$ A exciton (Fig. 3 (e)), the redistribution rate in Fig. 3 (g) indicates that scattered electrons have a wider distribution in the BZ than the MoS$_2$ A exciton, which means that electrons are able to move across the whole BZ and is consistent with the analysis in Fig. 2. For holes on the other hand we find that intravalley scattering is preferred. In the free electron-hole picture shown in Fig. 3 (h), we see that scattering of conduction electrons to a remote $\Lmin'$ has a higher intensity, which is a consequence of the stronger electron-phonon coupling between the $K$ and the $\Lmin'$ valley. Valence electrons scatter mostly to the side of the $\Gamma$ valley again due to energy conservation conditions. Comparing these two pictures, we can draw a conclusion that scattering in the exciton picture tends to redistribute charge over a wider range in the BZ, similar to the MoS$_2$ A exciton case. We suggest this is a result of the correlated nature of excitons.

\begin{figure}[t]
     \centering
     \includegraphics[width=.48\textwidth]{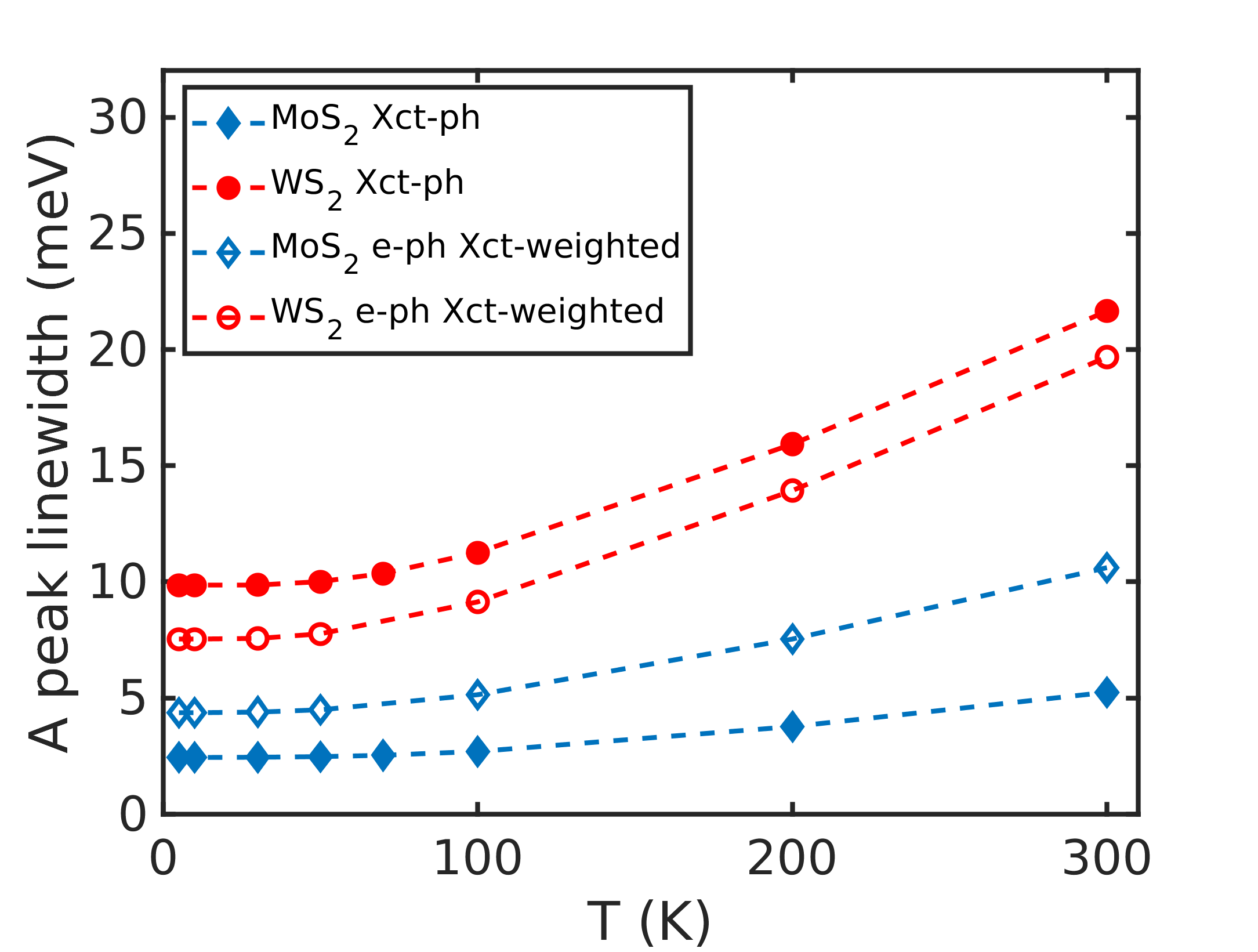}
\caption{Temperature dependent linewidth of MoS$_2$ (solid blue diamonds) and WS$_2$ (solid red dots) A excitons due to exciton-phonon couplings. Linewidth computed in the limit of zero exciton binding energy from the electron-phonon coupling weighted by exciton envelope functions are shown by the empty symbols. }
\label{fig:gammaT}
\end{figure}

The exciton-phonon coupling strength can be deduced experimentally by measuring the temperature-dependent exciton linewidth. In Fig.~\ref{fig:gammaT} we show the linewidths of both MoS$_2$ and WS$_2$ A excitons, which is computed by taking the on-shell imaginary part of the self-energy given by Eq.~\ref{Eq:sediag}. The exciton-phonon induced linewidth is about 5 meV and 22 meV at 300 K for MoS$_2$ and WS$_2$ A exciton, respectively. To explicitly see the effects of electron-hole interactions, we also compute the linewidth as a linear combination of electron-phonon (and hole-phonon) interaction weighted by the exciton envelope function, which we refer to as the exciton-weighted free electron-hole linewidth~\cite{Marini2008,Molina2016}. Although the exciton-phonon linewidth is within the same order of magnitude as that obtained from excited-weighted approach, there is no consistent relation between these two cases. For the MoS$_2$ A exciton, the simpler exciton-weighted electron-phonon linewidth shows a stronger temperature dependence and an overall larger linewidth, while the opposite is true for the for WS$_2$ A exciton. It is therefore necessary to compute exciton-phonon interactions directly for a correct description of exciton relaxations. Finally, we note that the linewidth of the MoS$_2$ A exciton increases to 5 meV in the MoS$_2$/WS$_2$ heterostructure compared to 2 meV in the monolayer at 300 K~\cite{Chan2023} while a 26 meV increase was reported in the experiment~\cite{Rigosi2015}. The underestimation might be owing to the fact that we do not consider defect scatterings and sample inhomogeneity in our calculations.

In conclusion, our first-principles calculations reveal the rich and ultrafast, phonon-mediated exciton scattering channels in a prototypical TMDC bilayer structure of WS$_2$/MoS$_2$. We show that the MoS$_2$ A exciton has a relaxation time of about 67 fs and WS$_2$ has a relaxation time of about 15 fs at 300 K. Moreover, we show that the ultrafast interlayer charge transfer takes place through a multiplicity of channels, and that two-step scattering processes play a significant role: for an initially excited A exciton in MoS$_2$, we predict such channels to cause a photobleaching of the absorption signal at the A exciton resonance in WS$_2$ in about 200 fs.
Our band BZ-resolved analysis further reveals that, upon excitation of the MoS$_2$ A exciton, the relaxation primarily involves exciton scatterings which transfer the hole from the $K$ point to the $\Gamma$ region of the BZ while, for the WS$_2$ A excition, it involves the scattering of the electron primarily to a valley around the $\Lambda$ valley. We expect these findings to inform novel ways of stacking, electronic hybridization, and many-body effects and be synergistically employed to tune charge and energy dynamics in TMDC heterostructures, and that the formalism described here may be used in future studies involving exciton transport in real-time.

Supporting Information. Computational details of GW-BSE, electron-phonon, and exciton-phonon couplings, supplementary figures of phonon dispersion, absorption spectrum, exciton envelope functions, estimation of bleaching time, electron-phonon coupling matrix elements, and derivation of the exciton population change rate.

This work was primarily supported by the Center for Computational Study of Excited State Phenomena in Energy Materials (C2SEPEM), which is funded by the U.S. Department of Energy, Office of Science, Basic Energy Sciences, Materials Sciences and Engineering Division under Contract No. DE-AC02-05CH11231, as part of the Computational Materials Sciences Program. FHJ acknowledges support from the National Science Foundation CAREER award through Grant No. DMR-2238328. Y.-H. C. was supported by the National Science and Technology Council of Taiwan under grant no. 110-2124-M-002-012. We acknowledge the use of computational resources at the National Energy Research Scientific Computing Center (NERSC), a DOE Office of Science User Facility supported by the Office of Science of the U.S. Department of Energy under Contract No. DE-AC02-05CH11231. The authors acknowledge the Texas Advanced Computing Center (TACC) at The University of Texas at Austin and National Center for High-performance Computing (NCHC) in Taiwan for providing HPC resources that have contributed to the research results reported within this paper.

%\begin{figure*}[h]
%     \centering
%     \includegraphics[width=\textwidth]{TOC}
%\caption{TOC image}
%\end{figure*}

\section*{References}
\bibliography{ref}

\begin{thebibliography}{40}%
\makeatletter
\providecommand \@ifxundefined [1]{%
 \@ifx{#1\undefined}
}%
\providecommand \@ifnum [1]{%
 \ifnum #1\expandafter \@firstoftwo
 \else \expandafter \@secondoftwo
 \fi
}%
\providecommand \@ifx [1]{%
 \ifx #1\expandafter \@firstoftwo
 \else \expandafter \@secondoftwo
 \fi
}%
\providecommand \natexlab [1]{#1}%
\providecommand \enquote  [1]{``#1''}%
\providecommand \bibnamefont  [1]{#1}%
\providecommand \bibfnamefont [1]{#1}%
\providecommand \citenamefont [1]{#1}%
\providecommand \href@noop [0]{\@secondoftwo}%
\providecommand \href [0]{\begingroup \@sanitize@url \@href}%
\providecommand \@href[1]{\@@startlink{#1}\@@href}%
\providecommand \@@href[1]{\endgroup#1\@@endlink}%
\providecommand \@sanitize@url [0]{\catcode `\\12\catcode `\$12\catcode
  `\&12\catcode `\#12\catcode `\^12\catcode `\_12\catcode `\%12\relax}%
\providecommand \@@startlink[1]{}%
\providecommand \@@endlink[0]{}%
\providecommand \url  [0]{\begingroup\@sanitize@url \@url }%
\providecommand \@url [1]{\endgroup\@href {#1}{\urlprefix }}%
\providecommand \urlprefix  [0]{URL }%
\providecommand \Eprint [0]{\href }%
\providecommand \doibase [0]{http://dx.doi.org/}%
\providecommand \selectlanguage [0]{\@gobble}%
\providecommand \bibinfo  [0]{\@secondoftwo}%
\providecommand \bibfield  [0]{\@secondoftwo}%
\providecommand \translation [1]{[#1]}%
\providecommand \BibitemOpen [0]{}%
\providecommand \bibitemStop [0]{}%
\providecommand \bibitemNoStop [0]{.\EOS\space}%
\providecommand \EOS [0]{\spacefactor3000\relax}%
\providecommand \BibitemShut  [1]{\csname bibitem#1\endcsname}%
\let\auto@bib@innerbib\@empty
%</preamble>
\bibitem [{\citenamefont {Geim}\ and\ \citenamefont
  {Grigorieva}(2013)}]{Geim2013}%
  \BibitemOpen
  \bibfield  {author} {\bibinfo {author} {\bibfnamefont {A.~K.}\ \bibnamefont
  {Geim}}\ and\ \bibinfo {author} {\bibfnamefont {I.~V.}\ \bibnamefont
  {Grigorieva}},\ }\bibfield  {title} {\enquote {\bibinfo {title} {Van der
  waals heterostructures},}\ }\href {https://doi.org/10.1038/nature12385}
  {\bibfield  {journal} {\bibinfo  {journal} {Nature}\ }\textbf {\bibinfo
  {volume} {499}},\ \bibinfo {pages} {419--425} (\bibinfo {year}
  {2013})}\BibitemShut {NoStop}%
\bibitem [{\citenamefont {Rivera}\ \emph {et~al.}(2018)\citenamefont {Rivera},
  \citenamefont {Yu}, \citenamefont {Seyler}, \citenamefont {Wilson},
  \citenamefont {Yao},\ and\ \citenamefont {Xu}}]{Rivera2018}%
  \BibitemOpen
  \bibfield  {author} {\bibinfo {author} {\bibfnamefont {Pasqual}\ \bibnamefont
  {Rivera}}, \bibinfo {author} {\bibfnamefont {Hongyi}\ \bibnamefont {Yu}},
  \bibinfo {author} {\bibfnamefont {Kyle~L.}\ \bibnamefont {Seyler}}, \bibinfo
  {author} {\bibfnamefont {Nathan~P.}\ \bibnamefont {Wilson}}, \bibinfo
  {author} {\bibfnamefont {Wang}\ \bibnamefont {Yao}}, \ and\ \bibinfo {author}
  {\bibfnamefont {Xiaodong}\ \bibnamefont {Xu}},\ }\bibfield  {title} {\enquote
  {\bibinfo {title} {Interlayer valley excitons in heterobilayers of transition
  metal dichalcogenides},}\ }\href {https://doi.org/10.1038/s41565-018-0193-0}
  {\bibfield  {journal} {\bibinfo  {journal} {Nature Nanotechnology}\ }\textbf
  {\bibinfo {volume} {13}},\ \bibinfo {pages} {1004--1015} (\bibinfo {year}
  {2018})}\BibitemShut {NoStop}%
\bibitem [{\citenamefont {Kim}\ \emph {et~al.}(2017)\citenamefont {Kim},
  \citenamefont {Jin}, \citenamefont {Chen}, \citenamefont {Cai}, \citenamefont
  {Zhao}, \citenamefont {Lee}, \citenamefont {Kahn}, \citenamefont {Watanabe},
  \citenamefont {Taniguchi}, \citenamefont {Tongay}, \citenamefont {Crommie},\
  and\ \citenamefont {Wang}}]{Kim2017}%
  \BibitemOpen
  \bibfield  {author} {\bibinfo {author} {\bibfnamefont {Jonghwan}\
  \bibnamefont {Kim}}, \bibinfo {author} {\bibfnamefont {Chenhao}\ \bibnamefont
  {Jin}}, \bibinfo {author} {\bibfnamefont {Bin}\ \bibnamefont {Chen}},
  \bibinfo {author} {\bibfnamefont {Hui}\ \bibnamefont {Cai}}, \bibinfo
  {author} {\bibfnamefont {Tao}\ \bibnamefont {Zhao}}, \bibinfo {author}
  {\bibfnamefont {Puiyee}\ \bibnamefont {Lee}}, \bibinfo {author}
  {\bibfnamefont {Salman}\ \bibnamefont {Kahn}}, \bibinfo {author}
  {\bibfnamefont {Kenji}\ \bibnamefont {Watanabe}}, \bibinfo {author}
  {\bibfnamefont {Takashi}\ \bibnamefont {Taniguchi}}, \bibinfo {author}
  {\bibfnamefont {Sefaattin}\ \bibnamefont {Tongay}}, \bibinfo {author}
  {\bibfnamefont {Michael~F.}\ \bibnamefont {Crommie}}, \ and\ \bibinfo
  {author} {\bibfnamefont {Feng}\ \bibnamefont {Wang}},\ }\bibfield  {title}
  {\enquote {\bibinfo {title} {Observation of ultralong valley lifetime in
  wse<sub>2</sub>/mos<sub>2</sub> heterostructures},}\ }\href {\doibase
  10.1126/sciadv.1700518} {\bibfield  {journal} {\bibinfo  {journal} {Science
  Advances}\ }\textbf {\bibinfo {volume} {3}},\ \bibinfo {pages} {e1700518}
  (\bibinfo {year} {2017})},\ \Eprint
  {http://arxiv.org/abs/https://www.science.org/doi/pdf/10.1126/sciadv.1700518}
  {https://www.science.org/doi/pdf/10.1126/sciadv.1700518} \BibitemShut
  {NoStop}%
\bibitem [{\citenamefont {Hong}\ \emph {et~al.}(2014)\citenamefont {Hong},
  \citenamefont {Kim}, \citenamefont {Shi}, \citenamefont {Zhang},
  \citenamefont {Jin}, \citenamefont {Sun}, \citenamefont {Tongay},
  \citenamefont {Wu}, \citenamefont {Zhang},\ and\ \citenamefont
  {Wang}}]{Hong2014}%
  \BibitemOpen
  \bibfield  {author} {\bibinfo {author} {\bibfnamefont {Xiaoping}\
  \bibnamefont {Hong}}, \bibinfo {author} {\bibfnamefont {Jonghwan}\
  \bibnamefont {Kim}}, \bibinfo {author} {\bibfnamefont {Su-Fei}\ \bibnamefont
  {Shi}}, \bibinfo {author} {\bibfnamefont {Yu}~\bibnamefont {Zhang}}, \bibinfo
  {author} {\bibfnamefont {Chenhao}\ \bibnamefont {Jin}}, \bibinfo {author}
  {\bibfnamefont {Yinghui}\ \bibnamefont {Sun}}, \bibinfo {author}
  {\bibfnamefont {Sefaattin}\ \bibnamefont {Tongay}}, \bibinfo {author}
  {\bibfnamefont {Junqiao}\ \bibnamefont {Wu}}, \bibinfo {author}
  {\bibfnamefont {Yanfeng}\ \bibnamefont {Zhang}}, \ and\ \bibinfo {author}
  {\bibfnamefont {Feng}\ \bibnamefont {Wang}},\ }\bibfield  {title} {\enquote
  {\bibinfo {title} {Ultrafast charge transfer in atomically thin mos2/ws2
  heterostructures},}\ }\href {https://doi.org/10.1038/nnano.2014.167}
  {\bibfield  {journal} {\bibinfo  {journal} {Nature Nanotechnology}\ }\textbf
  {\bibinfo {volume} {9}},\ \bibinfo {pages} {682--686} (\bibinfo {year}
  {2014})}\BibitemShut {NoStop}%
\bibitem [{\citenamefont {Yu}\ \emph {et~al.}(2015)\citenamefont {Yu},
  \citenamefont {Hu}, \citenamefont {Su}, \citenamefont {Huang}, \citenamefont
  {Liu}, \citenamefont {Jin}, \citenamefont {Purezky}, \citenamefont
  {Geohegan}, \citenamefont {Kim}, \citenamefont {Zhang},\ and\ \citenamefont
  {Cao}}]{Yu2015}%
  \BibitemOpen
  \bibfield  {author} {\bibinfo {author} {\bibfnamefont {Yifei}\ \bibnamefont
  {Yu}}, \bibinfo {author} {\bibfnamefont {Shi}\ \bibnamefont {Hu}}, \bibinfo
  {author} {\bibfnamefont {Liqin}\ \bibnamefont {Su}}, \bibinfo {author}
  {\bibfnamefont {Lujun}\ \bibnamefont {Huang}}, \bibinfo {author}
  {\bibfnamefont {Yi}~\bibnamefont {Liu}}, \bibinfo {author} {\bibfnamefont
  {Zhenghe}\ \bibnamefont {Jin}}, \bibinfo {author} {\bibfnamefont
  {Alexander~A.}\ \bibnamefont {Purezky}}, \bibinfo {author} {\bibfnamefont
  {David~B.}\ \bibnamefont {Geohegan}}, \bibinfo {author} {\bibfnamefont
  {Ki~Wook}\ \bibnamefont {Kim}}, \bibinfo {author} {\bibfnamefont {Yong}\
  \bibnamefont {Zhang}}, \ and\ \bibinfo {author} {\bibfnamefont {Linyou}\
  \bibnamefont {Cao}},\ }\bibfield  {title} {\enquote {\bibinfo {title}
  {Equally efficient interlayer exciton relaxation and improved absorption in
  epitaxial and nonepitaxial mos2/ws2 heterostructures},}\ }\href {\doibase
  10.1021/nl5038177} {\bibfield  {journal} {\bibinfo  {journal} {Nano Lett.}\
  }\textbf {\bibinfo {volume} {15}},\ \bibinfo {pages} {486--491} (\bibinfo
  {year} {2015})}\BibitemShut {NoStop}%
\bibitem [{\citenamefont {Ji}\ \emph {et~al.}(2017)\citenamefont {Ji},
  \citenamefont {Hong}, \citenamefont {Zhang}, \citenamefont {Zhang},
  \citenamefont {Huang}, \citenamefont {Cao}, \citenamefont {Qiao},
  \citenamefont {Liu}, \citenamefont {Liang}, \citenamefont {Jin},
  \citenamefont {Jiao}, \citenamefont {Shi}, \citenamefont {Meng},\ and\
  \citenamefont {Liu}}]{Ji2017}%
  \BibitemOpen
  \bibfield  {author} {\bibinfo {author} {\bibfnamefont {Ziheng}\ \bibnamefont
  {Ji}}, \bibinfo {author} {\bibfnamefont {Hao}\ \bibnamefont {Hong}}, \bibinfo
  {author} {\bibfnamefont {Jin}\ \bibnamefont {Zhang}}, \bibinfo {author}
  {\bibfnamefont {Qi}~\bibnamefont {Zhang}}, \bibinfo {author} {\bibfnamefont
  {Wei}\ \bibnamefont {Huang}}, \bibinfo {author} {\bibfnamefont {Ting}\
  \bibnamefont {Cao}}, \bibinfo {author} {\bibfnamefont {Ruixi}\ \bibnamefont
  {Qiao}}, \bibinfo {author} {\bibfnamefont {Can}\ \bibnamefont {Liu}},
  \bibinfo {author} {\bibfnamefont {Jing}\ \bibnamefont {Liang}}, \bibinfo
  {author} {\bibfnamefont {Chuanhong}\ \bibnamefont {Jin}}, \bibinfo {author}
  {\bibfnamefont {Liying}\ \bibnamefont {Jiao}}, \bibinfo {author}
  {\bibfnamefont {Kebin}\ \bibnamefont {Shi}}, \bibinfo {author} {\bibfnamefont
  {Sheng}\ \bibnamefont {Meng}}, \ and\ \bibinfo {author} {\bibfnamefont
  {Kaihui}\ \bibnamefont {Liu}},\ }\bibfield  {title} {\enquote {\bibinfo
  {title} {Robust stacking-independent ultrafast charge transfer in mos2/ws2
  bilayers},}\ }\href {\doibase 10.1021/acsnano.7b04541} {\bibfield  {journal}
  {\bibinfo  {journal} {ACS Nano}\ }\textbf {\bibinfo {volume} {11}},\ \bibinfo
  {pages} {12020--12026} (\bibinfo {year} {2017})}\BibitemShut {NoStop}%
\bibitem [{\citenamefont {Zhu}\ \emph {et~al.}(2017)\citenamefont {Zhu},
  \citenamefont {Wang}, \citenamefont {Gong}, \citenamefont {Kim},
  \citenamefont {Hone},\ and\ \citenamefont {Zhu}}]{Zhu2017}%
  \BibitemOpen
  \bibfield  {author} {\bibinfo {author} {\bibfnamefont {Haiming}\ \bibnamefont
  {Zhu}}, \bibinfo {author} {\bibfnamefont {Jue}\ \bibnamefont {Wang}},
  \bibinfo {author} {\bibfnamefont {Zizhou}\ \bibnamefont {Gong}}, \bibinfo
  {author} {\bibfnamefont {Young~Duck}\ \bibnamefont {Kim}}, \bibinfo {author}
  {\bibfnamefont {James}\ \bibnamefont {Hone}}, \ and\ \bibinfo {author}
  {\bibfnamefont {X.-Y.}\ \bibnamefont {Zhu}},\ }\bibfield  {title} {\enquote
  {\bibinfo {title} {Interfacial charge transfer circumventing momentum
  mismatch at two-dimensional van der waals heterojunctions},}\ }\href
  {\doibase 10.1021/acs.nanolett.7b00748} {\bibfield  {journal} {\bibinfo
  {journal} {Nano Lett.}\ }\textbf {\bibinfo {volume} {17}},\ \bibinfo {pages}
  {3591--3598} (\bibinfo {year} {2017})}\BibitemShut {NoStop}%
\bibitem [{\citenamefont {Jin}\ \emph {et~al.}(2018)\citenamefont {Jin},
  \citenamefont {Ma}, \citenamefont {Karni}, \citenamefont {Regan},
  \citenamefont {Wang},\ and\ \citenamefont {Heinz}}]{Jin2018}%
  \BibitemOpen
  \bibfield  {author} {\bibinfo {author} {\bibfnamefont {Chenhao}\ \bibnamefont
  {Jin}}, \bibinfo {author} {\bibfnamefont {Eric~Yue}\ \bibnamefont {Ma}},
  \bibinfo {author} {\bibfnamefont {Ouri}\ \bibnamefont {Karni}}, \bibinfo
  {author} {\bibfnamefont {Emma~C.}\ \bibnamefont {Regan}}, \bibinfo {author}
  {\bibfnamefont {Feng}\ \bibnamefont {Wang}}, \ and\ \bibinfo {author}
  {\bibfnamefont {Tony~F.}\ \bibnamefont {Heinz}},\ }\bibfield  {title}
  {\enquote {\bibinfo {title} {Ultrafast dynamics in van der waals
  heterostructures},}\ }\href {https://doi.org/10.1038/s41565-018-0298-5}
  {\bibfield  {journal} {\bibinfo  {journal} {Nature Nanotechnology}\ }\textbf
  {\bibinfo {volume} {13}},\ \bibinfo {pages} {994--1003} (\bibinfo {year}
  {2018})}\BibitemShut {NoStop}%
\bibitem [{\citenamefont {Zhou}\ \emph {et~al.}(2019)\citenamefont {Zhou},
  \citenamefont {Zhao},\ and\ \citenamefont {Zhu}}]{Zhou2019}%
  \BibitemOpen
  \bibfield  {author} {\bibinfo {author} {\bibfnamefont {Hongzhi}\ \bibnamefont
  {Zhou}}, \bibinfo {author} {\bibfnamefont {Yida}\ \bibnamefont {Zhao}}, \
  and\ \bibinfo {author} {\bibfnamefont {Haiming}\ \bibnamefont {Zhu}},\
  }\bibfield  {title} {\enquote {\bibinfo {title} {Dielectric
  environment-robust ultrafast charge transfer between two atomic layers},}\
  }\href {\doibase 10.1021/acs.jpclett.8b03596} {\bibfield  {journal} {\bibinfo
   {journal} {The Journal of Physical Chemistry Letters}\ }\textbf {\bibinfo
  {volume} {10}},\ \bibinfo {pages} {150--155} (\bibinfo {year} {2019})},\
  \Eprint {http://arxiv.org/abs/https://doi.org/10.1021/acs.jpclett.8b03596}
  {https://doi.org/10.1021/acs.jpclett.8b03596} \BibitemShut {NoStop}%
\bibitem [{\citenamefont {Rigosi}\ \emph {et~al.}(2015)\citenamefont {Rigosi},
  \citenamefont {Hill}, \citenamefont {Li}, \citenamefont {Chernikov},\ and\
  \citenamefont {Heinz}}]{Rigosi2015}%
  \BibitemOpen
  \bibfield  {author} {\bibinfo {author} {\bibfnamefont {Albert~F.}\
  \bibnamefont {Rigosi}}, \bibinfo {author} {\bibfnamefont {Heather~M.}\
  \bibnamefont {Hill}}, \bibinfo {author} {\bibfnamefont {Yilei}\ \bibnamefont
  {Li}}, \bibinfo {author} {\bibfnamefont {Alexey}\ \bibnamefont {Chernikov}},
  \ and\ \bibinfo {author} {\bibfnamefont {Tony~F.}\ \bibnamefont {Heinz}},\
  }\bibfield  {title} {\enquote {\bibinfo {title} {Probing interlayer
  interactions in transition metal dichalcogenide heterostructures by optical
  spectroscopy: Mos2/ws2 and mose2/wse2},}\ }\href {\doibase
  10.1021/acs.nanolett.5b01055} {\bibfield  {journal} {\bibinfo  {journal}
  {Nano Lett.}\ }\textbf {\bibinfo {volume} {15}},\ \bibinfo {pages}
  {5033--5038} (\bibinfo {year} {2015})}\BibitemShut {NoStop}%
\bibitem [{\citenamefont {Wang}\ \emph {et~al.}(2017)\citenamefont {Wang},
  \citenamefont {Wang}, \citenamefont {Yao}, \citenamefont {Liu},\ and\
  \citenamefont {Yu}}]{WangYong2017}%
  \BibitemOpen
  \bibfield  {author} {\bibinfo {author} {\bibfnamefont {Yong}\ \bibnamefont
  {Wang}}, \bibinfo {author} {\bibfnamefont {Zhan}\ \bibnamefont {Wang}},
  \bibinfo {author} {\bibfnamefont {Wang}\ \bibnamefont {Yao}}, \bibinfo
  {author} {\bibfnamefont {Gui-Bin}\ \bibnamefont {Liu}}, \ and\ \bibinfo
  {author} {\bibfnamefont {Hongyi}\ \bibnamefont {Yu}},\ }\bibfield  {title}
  {\enquote {\bibinfo {title} {Interlayer coupling in commensurate and
  incommensurate bilayer structures of transition-metal dichalcogenides},}\
  }\href {\doibase 10.1103/PhysRevB.95.115429} {\bibfield  {journal} {\bibinfo
  {journal} {Phys. Rev. B}\ }\textbf {\bibinfo {volume} {95}},\ \bibinfo
  {pages} {115429} (\bibinfo {year} {2017})}\BibitemShut {NoStop}%
\bibitem [{\citenamefont {Schaibley}\ \emph {et~al.}(2016)\citenamefont
  {Schaibley}, \citenamefont {Rivera}, \citenamefont {Yu}, \citenamefont
  {Seyler}, \citenamefont {Yan}, \citenamefont {Mandrus}, \citenamefont
  {Taniguchi}, \citenamefont {Watanabe}, \citenamefont {Yao},\ and\
  \citenamefont {Xu}}]{Schaibley2016}%
  \BibitemOpen
  \bibfield  {author} {\bibinfo {author} {\bibfnamefont {John~R.}\ \bibnamefont
  {Schaibley}}, \bibinfo {author} {\bibfnamefont {Pasqual}\ \bibnamefont
  {Rivera}}, \bibinfo {author} {\bibfnamefont {Hongyi}\ \bibnamefont {Yu}},
  \bibinfo {author} {\bibfnamefont {Kyle~L.}\ \bibnamefont {Seyler}}, \bibinfo
  {author} {\bibfnamefont {Jiaqiang}\ \bibnamefont {Yan}}, \bibinfo {author}
  {\bibfnamefont {David~G.}\ \bibnamefont {Mandrus}}, \bibinfo {author}
  {\bibfnamefont {Takashi}\ \bibnamefont {Taniguchi}}, \bibinfo {author}
  {\bibfnamefont {Kenji}\ \bibnamefont {Watanabe}}, \bibinfo {author}
  {\bibfnamefont {Wang}\ \bibnamefont {Yao}}, \ and\ \bibinfo {author}
  {\bibfnamefont {Xiaodong}\ \bibnamefont {Xu}},\ }\bibfield  {title} {\enquote
  {\bibinfo {title} {Directional interlayer spin-valley transfer in
  two-dimensional heterostructures},}\ }\href
  {https://doi.org/10.1038/ncomms13747} {\bibfield  {journal} {\bibinfo
  {journal} {Nature Communications}\ }\textbf {\bibinfo {volume} {7}},\
  \bibinfo {pages} {13747} (\bibinfo {year} {2016})}\BibitemShut {NoStop}%
\bibitem [{\citenamefont {Chen}\ \emph {et~al.}(2016)\citenamefont {Chen},
  \citenamefont {Wen}, \citenamefont {Zhang}, \citenamefont {Wu}, \citenamefont
  {Gong}, \citenamefont {Zhang}, \citenamefont {Yuan}, \citenamefont {Yi},
  \citenamefont {Lou}, \citenamefont {Ajayan}, \citenamefont {Zhuang},
  \citenamefont {Zhang},\ and\ \citenamefont {Zheng}}]{Chen2016}%
  \BibitemOpen
  \bibfield  {author} {\bibinfo {author} {\bibfnamefont {Hailong}\ \bibnamefont
  {Chen}}, \bibinfo {author} {\bibfnamefont {Xiewen}\ \bibnamefont {Wen}},
  \bibinfo {author} {\bibfnamefont {Jing}\ \bibnamefont {Zhang}}, \bibinfo
  {author} {\bibfnamefont {Tianmin}\ \bibnamefont {Wu}}, \bibinfo {author}
  {\bibfnamefont {Yongji}\ \bibnamefont {Gong}}, \bibinfo {author}
  {\bibfnamefont {Xiang}\ \bibnamefont {Zhang}}, \bibinfo {author}
  {\bibfnamefont {Jiangtan}\ \bibnamefont {Yuan}}, \bibinfo {author}
  {\bibfnamefont {Chongyue}\ \bibnamefont {Yi}}, \bibinfo {author}
  {\bibfnamefont {Jun}\ \bibnamefont {Lou}}, \bibinfo {author} {\bibfnamefont
  {Pulickel~M.}\ \bibnamefont {Ajayan}}, \bibinfo {author} {\bibfnamefont
  {Wei}\ \bibnamefont {Zhuang}}, \bibinfo {author} {\bibfnamefont {Guangyu}\
  \bibnamefont {Zhang}}, \ and\ \bibinfo {author} {\bibfnamefont {Junrong}\
  \bibnamefont {Zheng}},\ }\bibfield  {title} {\enquote {\bibinfo {title}
  {Ultrafast formation of interlayer hot excitons in atomically thin mos2/ws2
  heterostructures},}\ }\href {https://doi.org/10.1038/ncomms12512} {\bibfield
  {journal} {\bibinfo  {journal} {Nature Communications}\ }\textbf {\bibinfo
  {volume} {7}},\ \bibinfo {pages} {12512} (\bibinfo {year}
  {2016})}\BibitemShut {NoStop}%
\bibitem [{\citenamefont {Nagler}\ \emph {et~al.}(2017)\citenamefont {Nagler},
  \citenamefont {Plechinger}, \citenamefont {Ballottin}, \citenamefont
  {Mitioglu}, \citenamefont {Meier}, \citenamefont {Paradiso}, \citenamefont
  {Strunk}, \citenamefont {Chernikov}, \citenamefont {Christianen},
  \citenamefont {Schüller},\ and\ \citenamefont {Korn}}]{Nagler2017}%
  \BibitemOpen
  \bibfield  {author} {\bibinfo {author} {\bibfnamefont {Philipp}\ \bibnamefont
  {Nagler}}, \bibinfo {author} {\bibfnamefont {Gerd}\ \bibnamefont
  {Plechinger}}, \bibinfo {author} {\bibfnamefont {Mariana~V}\ \bibnamefont
  {Ballottin}}, \bibinfo {author} {\bibfnamefont {Anatolie}\ \bibnamefont
  {Mitioglu}}, \bibinfo {author} {\bibfnamefont {Sebastian}\ \bibnamefont
  {Meier}}, \bibinfo {author} {\bibfnamefont {Nicola}\ \bibnamefont
  {Paradiso}}, \bibinfo {author} {\bibfnamefont {Christoph}\ \bibnamefont
  {Strunk}}, \bibinfo {author} {\bibfnamefont {Alexey}\ \bibnamefont
  {Chernikov}}, \bibinfo {author} {\bibfnamefont {Peter C~M}\ \bibnamefont
  {Christianen}}, \bibinfo {author} {\bibfnamefont {Christian}\ \bibnamefont
  {Schüller}}, \ and\ \bibinfo {author} {\bibfnamefont {Tobias}\ \bibnamefont
  {Korn}},\ }\bibfield  {title} {\enquote {\bibinfo {title} {Interlayer exciton
  dynamics in a dichalcogenide monolayer heterostructure},}\ }\href {\doibase
  10.1088/2053-1583/aa7352} {\ \textbf {\bibinfo {volume} {4}},\ \bibinfo
  {pages} {025112} (\bibinfo {year} {2017})}\BibitemShut {NoStop}%
\bibitem [{\citenamefont {Bian}\ \emph {et~al.}(2020)\citenamefont {Bian},
  \citenamefont {He}, \citenamefont {Hao}, \citenamefont {Fu}, \citenamefont
  {Zhang}, \citenamefont {He}, \citenamefont {Wang},\ and\ \citenamefont
  {Zhao}}]{Bian2020}%
  \BibitemOpen
  \bibfield  {author} {\bibinfo {author} {\bibfnamefont {Ang}\ \bibnamefont
  {Bian}}, \bibinfo {author} {\bibfnamefont {Dawei}\ \bibnamefont {He}},
  \bibinfo {author} {\bibfnamefont {Shengcai}\ \bibnamefont {Hao}}, \bibinfo
  {author} {\bibfnamefont {Yang}\ \bibnamefont {Fu}}, \bibinfo {author}
  {\bibfnamefont {Lu}~\bibnamefont {Zhang}}, \bibinfo {author} {\bibfnamefont
  {Jiaqi}\ \bibnamefont {He}}, \bibinfo {author} {\bibfnamefont {Yongsheng}\
  \bibnamefont {Wang}}, \ and\ \bibinfo {author} {\bibfnamefont {Hui}\
  \bibnamefont {Zhao}},\ }\bibfield  {title} {\enquote {\bibinfo {title}
  {Dynamics of charge-transfer excitons in a transition metal dichalcogenide
  heterostructure},}\ }\href {\doibase 10.1039/D0NR01924K} {\bibfield
  {journal} {\bibinfo  {journal} {Nanoscale}\ }\textbf {\bibinfo {volume}
  {12}},\ \bibinfo {pages} {8485--8492} (\bibinfo {year} {2020})}\BibitemShut
  {NoStop}%
\bibitem [{\citenamefont {Policht}\ \emph {et~al.}(2021)\citenamefont
  {Policht}, \citenamefont {Russo}, \citenamefont {Liu}, \citenamefont
  {Trovatello}, \citenamefont {Maiuri}, \citenamefont {Bai}, \citenamefont
  {Zhu}, \citenamefont {Dal~Conte},\ and\ \citenamefont
  {Cerullo}}]{Policht2021}%
  \BibitemOpen
  \bibfield  {author} {\bibinfo {author} {\bibfnamefont {Veronica~R.}\
  \bibnamefont {Policht}}, \bibinfo {author} {\bibfnamefont {Mattia}\
  \bibnamefont {Russo}}, \bibinfo {author} {\bibfnamefont {Fang}\ \bibnamefont
  {Liu}}, \bibinfo {author} {\bibfnamefont {Chiara}\ \bibnamefont
  {Trovatello}}, \bibinfo {author} {\bibfnamefont {Margherita}\ \bibnamefont
  {Maiuri}}, \bibinfo {author} {\bibfnamefont {Yusong}\ \bibnamefont {Bai}},
  \bibinfo {author} {\bibfnamefont {Xiaoyang}\ \bibnamefont {Zhu}}, \bibinfo
  {author} {\bibfnamefont {Stefano}\ \bibnamefont {Dal~Conte}}, \ and\ \bibinfo
  {author} {\bibfnamefont {Giulio}\ \bibnamefont {Cerullo}},\ }\bibfield
  {title} {\enquote {\bibinfo {title} {Dissecting interlayer hole and electron
  transfer in transition metal dichalcogenide heterostructures via
  two-dimensional electronic spectroscopy},}\ }\href {\doibase
  10.1021/acs.nanolett.1c01098} {\bibfield  {journal} {\bibinfo  {journal}
  {Nano Lett.}\ }\textbf {\bibinfo {volume} {21}},\ \bibinfo {pages}
  {4738--4743} (\bibinfo {year} {2021})}\BibitemShut {NoStop}%
\bibitem [{\citenamefont {Wang}\ \emph {et~al.}(2016)\citenamefont {Wang},
  \citenamefont {Bang}, \citenamefont {Sun}, \citenamefont {Liang},
  \citenamefont {West}, \citenamefont {Meunier},\ and\ \citenamefont
  {Zhang}}]{Wang2016}%
  \BibitemOpen
  \bibfield  {author} {\bibinfo {author} {\bibfnamefont {Han}\ \bibnamefont
  {Wang}}, \bibinfo {author} {\bibfnamefont {Junhyeok}\ \bibnamefont {Bang}},
  \bibinfo {author} {\bibfnamefont {Yiyang}\ \bibnamefont {Sun}}, \bibinfo
  {author} {\bibfnamefont {Liangbo}\ \bibnamefont {Liang}}, \bibinfo {author}
  {\bibfnamefont {Damien}\ \bibnamefont {West}}, \bibinfo {author}
  {\bibfnamefont {Vincent}\ \bibnamefont {Meunier}}, \ and\ \bibinfo {author}
  {\bibfnamefont {Shengbai}\ \bibnamefont {Zhang}},\ }\bibfield  {title}
  {\enquote {\bibinfo {title} {The role of collective motion in the ultrafast
  charge transfer in van der waals heterostructures},}\ }\href
  {https://doi.org/10.1038/ncomms11504} {\bibfield  {journal} {\bibinfo
  {journal} {Nature Communications}\ }\textbf {\bibinfo {volume} {7}},\
  \bibinfo {pages} {11504} (\bibinfo {year} {2016})}\BibitemShut {NoStop}%
\bibitem [{\citenamefont {Long}\ and\ \citenamefont
  {Prezhdo}(2016)}]{Long2016}%
  \BibitemOpen
  \bibfield  {author} {\bibinfo {author} {\bibfnamefont {Run}\ \bibnamefont
  {Long}}\ and\ \bibinfo {author} {\bibfnamefont {Oleg~V.}\ \bibnamefont
  {Prezhdo}},\ }\bibfield  {title} {\enquote {\bibinfo {title} {Quantum
  coherence facilitates efficient charge separation at a mos2/mose2 van der
  waals junction},}\ }\href {\doibase 10.1021/acs.nanolett.5b05264} {\bibfield
  {journal} {\bibinfo  {journal} {Nano Lett.}\ }\textbf {\bibinfo {volume}
  {16}},\ \bibinfo {pages} {1996--2003} (\bibinfo {year} {2016})}\BibitemShut
  {NoStop}%
\bibitem [{\citenamefont {Zheng}\ \emph {et~al.}(2018)\citenamefont {Zheng},
  \citenamefont {Xie}, \citenamefont {Lan}, \citenamefont {Prezhdo},
  \citenamefont {Saidi},\ and\ \citenamefont {Zhao}}]{Zheng2017}%
  \BibitemOpen
  \bibfield  {author} {\bibinfo {author} {\bibfnamefont {Qijing}\ \bibnamefont
  {Zheng}}, \bibinfo {author} {\bibfnamefont {Yu}~\bibnamefont {Xie}}, \bibinfo
  {author} {\bibfnamefont {Zhenggang}\ \bibnamefont {Lan}}, \bibinfo {author}
  {\bibfnamefont {Oleg~V.}\ \bibnamefont {Prezhdo}}, \bibinfo {author}
  {\bibfnamefont {Wissam~A.}\ \bibnamefont {Saidi}}, \ and\ \bibinfo {author}
  {\bibfnamefont {Jin}\ \bibnamefont {Zhao}},\ }\bibfield  {title} {\enquote
  {\bibinfo {title} {Phonon-coupled ultrafast interlayer charge oscillation at
  van der waals heterostructure interfaces},}\ }\href {\doibase
  10.1103/PhysRevB.97.205417} {\bibfield  {journal} {\bibinfo  {journal} {Phys.
  Rev. B}\ }\textbf {\bibinfo {volume} {97}},\ \bibinfo {pages} {205417}
  (\bibinfo {year} {2018})}\BibitemShut {NoStop}%
\bibitem [{\citenamefont {Zheng}\ \emph {et~al.}(2017)\citenamefont {Zheng},
  \citenamefont {Saidi}, \citenamefont {Xie}, \citenamefont {Lan},
  \citenamefont {Prezhdo}, \citenamefont {Petek},\ and\ \citenamefont
  {Zhao}}]{Zheng2017nano}%
  \BibitemOpen
  \bibfield  {author} {\bibinfo {author} {\bibfnamefont {Qijing}\ \bibnamefont
  {Zheng}}, \bibinfo {author} {\bibfnamefont {Wissam~A.}\ \bibnamefont
  {Saidi}}, \bibinfo {author} {\bibfnamefont {Yu}~\bibnamefont {Xie}}, \bibinfo
  {author} {\bibfnamefont {Zhenggang}\ \bibnamefont {Lan}}, \bibinfo {author}
  {\bibfnamefont {Oleg~V.}\ \bibnamefont {Prezhdo}}, \bibinfo {author}
  {\bibfnamefont {Hrvoje}\ \bibnamefont {Petek}}, \ and\ \bibinfo {author}
  {\bibfnamefont {Jin}\ \bibnamefont {Zhao}},\ }\bibfield  {title} {\enquote
  {\bibinfo {title} {Phonon-assisted ultrafast charge transfer at van der waals
  heterostructure interface},}\ }\href {\doibase 10.1021/acs.nanolett.7b03429}
  {\bibfield  {journal} {\bibinfo  {journal} {Nano Lett.}\ }\textbf {\bibinfo
  {volume} {17}},\ \bibinfo {pages} {6435--6442} (\bibinfo {year}
  {2017})}\BibitemShut {NoStop}%
\bibitem [{\citenamefont {Zhang}\ \emph {et~al.}(2017)\citenamefont {Zhang},
  \citenamefont {Hong}, \citenamefont {Lian}, \citenamefont {Ma}, \citenamefont
  {Xu}, \citenamefont {Zhou}, \citenamefont {Fu}, \citenamefont {Liu},\ and\
  \citenamefont {Meng}}]{ZhangMeng2017}%
  \BibitemOpen
  \bibfield  {author} {\bibinfo {author} {\bibfnamefont {Jin}\ \bibnamefont
  {Zhang}}, \bibinfo {author} {\bibfnamefont {Hao}\ \bibnamefont {Hong}},
  \bibinfo {author} {\bibfnamefont {Chao}\ \bibnamefont {Lian}}, \bibinfo
  {author} {\bibfnamefont {Wei}\ \bibnamefont {Ma}}, \bibinfo {author}
  {\bibfnamefont {Xiaozhi}\ \bibnamefont {Xu}}, \bibinfo {author}
  {\bibfnamefont {Xu}~\bibnamefont {Zhou}}, \bibinfo {author} {\bibfnamefont
  {Huixia}\ \bibnamefont {Fu}}, \bibinfo {author} {\bibfnamefont {Kaihui}\
  \bibnamefont {Liu}}, \ and\ \bibinfo {author} {\bibfnamefont {Sheng}\
  \bibnamefont {Meng}},\ }\bibfield  {title} {\enquote {\bibinfo {title}
  {Interlayer-state-coupling dependent ultrafast charge transfer in mos2/ws2
  bilayers},}\ }\href {\doibase https://doi.org/10.1002/advs.201700086}
  {\bibfield  {journal} {\bibinfo  {journal} {Advanced Science}\ }\textbf
  {\bibinfo {volume} {4}},\ \bibinfo {pages} {1700086} (\bibinfo {year}
  {2017})},\ \Eprint
  {http://arxiv.org/abs/https://onlinelibrary.wiley.com/doi/pdf/10.1002/advs.201700086}
  {https://onlinelibrary.wiley.com/doi/pdf/10.1002/advs.201700086} \BibitemShut
  {NoStop}%
\bibitem [{\citenamefont {Wang}\ \emph {et~al.}(2021)\citenamefont {Wang},
  \citenamefont {Altmann}, \citenamefont {Gadermaier}, \citenamefont {Yang},
  \citenamefont {Li}, \citenamefont {Ghirardini}, \citenamefont {Trovatello},
  \citenamefont {Finazzi}, \citenamefont {Duò}, \citenamefont {Celebrano},
  \citenamefont {Long}, \citenamefont {Akinwande}, \citenamefont {Prezhdo},
  \citenamefont {Cerullo},\ and\ \citenamefont {Dal~Conte}}]{Wang2021}%
  \BibitemOpen
  \bibfield  {author} {\bibinfo {author} {\bibfnamefont {Zilong}\ \bibnamefont
  {Wang}}, \bibinfo {author} {\bibfnamefont {Patrick}\ \bibnamefont {Altmann}},
  \bibinfo {author} {\bibfnamefont {Christoph}\ \bibnamefont {Gadermaier}},
  \bibinfo {author} {\bibfnamefont {Yating}\ \bibnamefont {Yang}}, \bibinfo
  {author} {\bibfnamefont {Wei}\ \bibnamefont {Li}}, \bibinfo {author}
  {\bibfnamefont {Lavinia}\ \bibnamefont {Ghirardini}}, \bibinfo {author}
  {\bibfnamefont {Chiara}\ \bibnamefont {Trovatello}}, \bibinfo {author}
  {\bibfnamefont {Marco}\ \bibnamefont {Finazzi}}, \bibinfo {author}
  {\bibfnamefont {Lamberto}\ \bibnamefont {Duò}}, \bibinfo {author}
  {\bibfnamefont {Michele}\ \bibnamefont {Celebrano}}, \bibinfo {author}
  {\bibfnamefont {Run}\ \bibnamefont {Long}}, \bibinfo {author} {\bibfnamefont
  {Deji}\ \bibnamefont {Akinwande}}, \bibinfo {author} {\bibfnamefont
  {Oleg~V.}\ \bibnamefont {Prezhdo}}, \bibinfo {author} {\bibfnamefont
  {Giulio}\ \bibnamefont {Cerullo}}, \ and\ \bibinfo {author} {\bibfnamefont
  {Stefano}\ \bibnamefont {Dal~Conte}},\ }\bibfield  {title} {\enquote
  {\bibinfo {title} {Phonon-mediated interlayer charge separation and
  recombination in a mose2/wse2 heterostructure},}\ }\href {\doibase
  10.1021/acs.nanolett.0c04955} {\bibfield  {journal} {\bibinfo  {journal}
  {Nano Letters}\ }\textbf {\bibinfo {volume} {21}},\ \bibinfo {pages}
  {2165--2173} (\bibinfo {year} {2021})},\ \bibinfo {note} {pMID: 33591207},\
  \Eprint {http://arxiv.org/abs/https://doi.org/10.1021/acs.nanolett.0c04955}
  {https://doi.org/10.1021/acs.nanolett.0c04955} \BibitemShut {NoStop}%
\bibitem [{\citenamefont {Zeng}\ \emph {et~al.}(2021)\citenamefont {Zeng},
  \citenamefont {Liu}, \citenamefont {Zhang},\ and\ \citenamefont
  {Cheng}}]{Zeng2021}%
  \BibitemOpen
  \bibfield  {author} {\bibinfo {author} {\bibfnamefont {Huadong}\ \bibnamefont
  {Zeng}}, \bibinfo {author} {\bibfnamefont {Xiangyue}\ \bibnamefont {Liu}},
  \bibinfo {author} {\bibfnamefont {Hong}\ \bibnamefont {Zhang}}, \ and\
  \bibinfo {author} {\bibfnamefont {Xinlu}\ \bibnamefont {Cheng}},\ }\bibfield
  {title} {\enquote {\bibinfo {title} {New theoretical insights into the
  photoinduced carrier transfer dynamics in ws2/wse2 van der waals
  heterostructures},}\ }\href {\doibase 10.1039/D0CP04517A} {\bibfield
  {journal} {\bibinfo  {journal} {Phys. Chem. Chem. Phys.}\ }\textbf {\bibinfo
  {volume} {23}},\ \bibinfo {pages} {694--701} (\bibinfo {year}
  {2021})}\BibitemShut {NoStop}%
\bibitem [{\citenamefont {Liu}\ \emph {et~al.}(2020)\citenamefont {Liu},
  \citenamefont {Zhang},\ and\ \citenamefont {Lu}}]{Liu2020}%
  \BibitemOpen
  \bibfield  {author} {\bibinfo {author} {\bibfnamefont {Junyi}\ \bibnamefont
  {Liu}}, \bibinfo {author} {\bibfnamefont {Xu}~\bibnamefont {Zhang}}, \ and\
  \bibinfo {author} {\bibfnamefont {Gang}\ \bibnamefont {Lu}},\ }\bibfield
  {title} {\enquote {\bibinfo {title} {Excitonic effect drives ultrafast
  dynamics in van der waals heterostructures},}\ }\href {\doibase
  10.1021/acs.nanolett.0c01519} {\bibfield  {journal} {\bibinfo  {journal}
  {Nano Lett.}\ }\textbf {\bibinfo {volume} {20}},\ \bibinfo {pages}
  {4631--4637} (\bibinfo {year} {2020})}\BibitemShut {NoStop}%
\bibitem [{\citenamefont {Jiang}\ \emph {et~al.}(2021)\citenamefont {Jiang},
  \citenamefont {Zheng}, \citenamefont {Lan}, \citenamefont {Saidi},
  \citenamefont {Ren},\ and\ \citenamefont {Zhao}}]{Jian2021}%
  \BibitemOpen
  \bibfield  {author} {\bibinfo {author} {\bibfnamefont {Xiang}\ \bibnamefont
  {Jiang}}, \bibinfo {author} {\bibfnamefont {Qijing}\ \bibnamefont {Zheng}},
  \bibinfo {author} {\bibfnamefont {Zhenggang}\ \bibnamefont {Lan}}, \bibinfo
  {author} {\bibfnamefont {Wissam~A.}\ \bibnamefont {Saidi}}, \bibinfo {author}
  {\bibfnamefont {Xinguo}\ \bibnamefont {Ren}}, \ and\ \bibinfo {author}
  {\bibfnamefont {Jin}\ \bibnamefont {Zhao}},\ }\bibfield  {title} {\enquote
  {\bibinfo {title} {Real-time <i>gw</i>-bse investigations on spin-valley
  exciton dynamics in monolayer transition metal dichalcogenide},}\ }\href
  {\doibase 10.1126/sciadv.abf3759} {\bibfield  {journal} {\bibinfo  {journal}
  {Science Advances}\ }\textbf {\bibinfo {volume} {7}},\ \bibinfo {pages}
  {eabf3759} (\bibinfo {year} {2021})},\ \Eprint
  {http://arxiv.org/abs/https://www.science.org/doi/pdf/10.1126/sciadv.abf3759}
  {https://www.science.org/doi/pdf/10.1126/sciadv.abf3759} \BibitemShut
  {NoStop}%
\bibitem [{\citenamefont {Hu}\ \emph {et~al.}(2023)\citenamefont {Hu},
  \citenamefont {Naik}, \citenamefont {Chan},\ and\ \citenamefont
  {Louie}}]{Chen2023}%
  \BibitemOpen
  \bibfield  {author} {\bibinfo {author} {\bibfnamefont {Chen}\ \bibnamefont
  {Hu}}, \bibinfo {author} {\bibfnamefont {Mit~H.}\ \bibnamefont {Naik}},
  \bibinfo {author} {\bibfnamefont {Yang-hao}\ \bibnamefont {Chan}}, \ and\
  \bibinfo {author} {\bibfnamefont {Steven~G.}\ \bibnamefont {Louie}},\
  }\bibfield  {title} {\enquote {\bibinfo {title} {Excitonic interactions and
  mechanism for ultrafast interlayer photoexcited response in van der waals
  heterostructures},}\ }\href@noop {} {\bibfield  {journal} {\bibinfo
  {journal} {arXiv:2305.17335}\ } (\bibinfo {year} {2023})}\BibitemShut
  {NoStop}%
\bibitem [{\citenamefont {Qiu}\ \emph {et~al.}(2016)\citenamefont {Qiu},
  \citenamefont {da~Jornada},\ and\ \citenamefont {Louie}}]{Qiu2016}%
  \BibitemOpen
  \bibfield  {author} {\bibinfo {author} {\bibfnamefont {Diana~Y.}\
  \bibnamefont {Qiu}}, \bibinfo {author} {\bibfnamefont {Felipe~H.}\
  \bibnamefont {da~Jornada}}, \ and\ \bibinfo {author} {\bibfnamefont
  {Steven~G.}\ \bibnamefont {Louie}},\ }\bibfield  {title} {\enquote {\bibinfo
  {title} {Screening and many-body effects in two-dimensional crystals:
  Monolayer mos$_2$},}\ }\href@noop {} {\bibfield  {journal} {\bibinfo
  {journal} {Phys. Rev. B}\ }\textbf {\bibinfo {volume} {93}},\ \bibinfo
  {pages} {235435} (\bibinfo {year} {2016})}\BibitemShut {NoStop}%
\bibitem [{\citenamefont {Antonius}\ and\ \citenamefont
  {Louie}(2022)}]{Antonius2017}%
  \BibitemOpen
  \bibfield  {author} {\bibinfo {author} {\bibfnamefont {Gabriel}\ \bibnamefont
  {Antonius}}\ and\ \bibinfo {author} {\bibfnamefont {Steven~G.}\ \bibnamefont
  {Louie}},\ }\bibfield  {title} {\enquote {\bibinfo {title} {Theory of
  exciton-phonon coupling},}\ }\href {\doibase 10.1103/PhysRevB.105.085111}
  {\bibfield  {journal} {\bibinfo  {journal} {Phys. Rev. B}\ }\textbf {\bibinfo
  {volume} {105}},\ \bibinfo {pages} {085111} (\bibinfo {year}
  {2022})}\BibitemShut {NoStop}%
\bibitem [{\citenamefont {Chen}\ \emph {et~al.}(2020)\citenamefont {Chen},
  \citenamefont {Sangalli},\ and\ \citenamefont {Bernardi}}]{Chen2020}%
  \BibitemOpen
  \bibfield  {author} {\bibinfo {author} {\bibfnamefont {Hsiao-Yi}\
  \bibnamefont {Chen}}, \bibinfo {author} {\bibfnamefont {Davide}\ \bibnamefont
  {Sangalli}}, \ and\ \bibinfo {author} {\bibfnamefont {Marco}\ \bibnamefont
  {Bernardi}},\ }\bibfield  {title} {\enquote {\bibinfo {title} {Exciton-phonon
  interaction and relaxation times from first principles},}\ }\href {\doibase
  10.1103/PhysRevLett.125.107401} {\bibfield  {journal} {\bibinfo  {journal}
  {Phys. Rev. Lett.}\ }\textbf {\bibinfo {volume} {125}},\ \bibinfo {pages}
  {107401} (\bibinfo {year} {2020})}\BibitemShut {NoStop}%
\bibitem [{\citenamefont {Huang}\ \emph {et~al.}(2021)\citenamefont {Huang},
  \citenamefont {Zacharias}, \citenamefont {Lewis}, \citenamefont {Giustino},\
  and\ \citenamefont {Sharifzadeh}}]{huang2021}%
  \BibitemOpen
  \bibfield  {author} {\bibinfo {author} {\bibfnamefont {Tianlun~Allan}\
  \bibnamefont {Huang}}, \bibinfo {author} {\bibfnamefont {Marios}\
  \bibnamefont {Zacharias}}, \bibinfo {author} {\bibfnamefont {D.~Kirk}\
  \bibnamefont {Lewis}}, \bibinfo {author} {\bibfnamefont {Feliciano}\
  \bibnamefont {Giustino}}, \ and\ \bibinfo {author} {\bibfnamefont {Sahar}\
  \bibnamefont {Sharifzadeh}},\ }\bibfield  {title} {\enquote {\bibinfo {title}
  {Exciton–{Phonon} {Interactions} in {Monolayer} {Germanium} {Selenide} from
  {First} {Principles}},}\ }\href {\doibase 10.1021/acs.jpclett.1c00264}
  {\bibfield  {journal} {\bibinfo  {journal} {The Journal of Physical Chemistry
  Letters}\ }\textbf {\bibinfo {volume} {12}},\ \bibinfo {pages} {3802--3808}
  (\bibinfo {year} {2021})},\ \bibinfo {note} {publisher: American Chemical
  Society}\BibitemShut {NoStop}%
\bibitem [{\citenamefont {Hybertsen}\ and\ \citenamefont
  {Louie}(1986)}]{Hybertsen1986}%
  \BibitemOpen
  \bibfield  {author} {\bibinfo {author} {\bibfnamefont {Mark~S.}\ \bibnamefont
  {Hybertsen}}\ and\ \bibinfo {author} {\bibfnamefont {Steven~G.}\ \bibnamefont
  {Louie}},\ }\bibfield  {title} {\enquote {\bibinfo {title} {Electron
  correlation in semiconductors and insulators: Band gaps and quasiparticle
  energies},}\ }\href {\doibase 10.1103/PhysRevB.34.5390} {\bibfield  {journal}
  {\bibinfo  {journal} {Phys. Rev. B}\ }\textbf {\bibinfo {volume} {34}},\
  \bibinfo {pages} {5390--5413} (\bibinfo {year} {1986})}\BibitemShut {NoStop}%
\bibitem [{\citenamefont {Deslippe}\ \emph {et~al.}(2012)\citenamefont
  {Deslippe}, \citenamefont {Samsonidze}, \citenamefont {Strubbe},
  \citenamefont {Jain}, \citenamefont {Cohen},\ and\ \citenamefont
  {Louie}}]{Deslippe2012}%
  \BibitemOpen
  \bibfield  {author} {\bibinfo {author} {\bibfnamefont {Jack}\ \bibnamefont
  {Deslippe}}, \bibinfo {author} {\bibfnamefont {Georgy}\ \bibnamefont
  {Samsonidze}}, \bibinfo {author} {\bibfnamefont {David~A.}\ \bibnamefont
  {Strubbe}}, \bibinfo {author} {\bibfnamefont {Manish}\ \bibnamefont {Jain}},
  \bibinfo {author} {\bibfnamefont {Marvin~L.}\ \bibnamefont {Cohen}}, \ and\
  \bibinfo {author} {\bibfnamefont {Steven~G.}\ \bibnamefont {Louie}},\
  }\bibfield  {title} {\enquote {\bibinfo {title} {Berkeleygw: A massively
  parallel computer package for the calculation of the quasiparticle and
  optical properties of materials and nanostructures},}\ }\href
  {http://www.sciencedirect.com/science/article/pii/S0010465511003912}
  {\bibfield  {journal} {\bibinfo  {journal} {Computer Physics Communications}\
  }\textbf {\bibinfo {volume} {183}},\ \bibinfo {pages} {1269--1289} (\bibinfo
  {year} {2012})}\BibitemShut {NoStop}%
\bibitem [{\citenamefont {Rohlfing}\ and\ \citenamefont
  {Louie}(2000)}]{Rohlfing2000}%
  \BibitemOpen
  \bibfield  {author} {\bibinfo {author} {\bibfnamefont {Michael}\ \bibnamefont
  {Rohlfing}}\ and\ \bibinfo {author} {\bibfnamefont {Steven~G.}\ \bibnamefont
  {Louie}},\ }\bibfield  {title} {\enquote {\bibinfo {title} {Electron-hole
  excitations and optical spectra from first principles},}\ }\href {\doibase
  10.1103/PhysRevB.62.4927} {\bibfield  {journal} {\bibinfo  {journal} {Phys.
  Rev. B}\ }\textbf {\bibinfo {volume} {62}},\ \bibinfo {pages} {4927--4944}
  (\bibinfo {year} {2000})}\BibitemShut {NoStop}%
\bibitem [{\citenamefont {Qiu}\ \emph {et~al.}(2015)\citenamefont {Qiu},
  \citenamefont {Cao},\ and\ \citenamefont {Louie}}]{Qiu2015}%
  \BibitemOpen
  \bibfield  {author} {\bibinfo {author} {\bibfnamefont {Diana~Y.}\
  \bibnamefont {Qiu}}, \bibinfo {author} {\bibfnamefont {Ting}\ \bibnamefont
  {Cao}}, \ and\ \bibinfo {author} {\bibfnamefont {Steven~G.}\ \bibnamefont
  {Louie}},\ }\bibfield  {title} {\enquote {\bibinfo {title} {Nonanalyticity,
  valley quantum phases, and lightlike exciton dispersion in monolayer
  transition metal dichalcogenides: Theory and first-principles
  calculations},}\ }\href {\doibase 10.1103/PhysRevLett.115.176801} {\bibfield
  {journal} {\bibinfo  {journal} {Phys. Rev. Lett.}\ }\textbf {\bibinfo
  {volume} {115}},\ \bibinfo {pages} {176801} (\bibinfo {year}
  {2015})}\BibitemShut {NoStop}%
\bibitem [{\citenamefont {Torun}\ \emph {et~al.}(2018)\citenamefont {Torun},
  \citenamefont {Miranda}, \citenamefont {Molina-S\'anchez},\ and\
  \citenamefont {Wirtz}}]{Torun2018}%
  \BibitemOpen
  \bibfield  {author} {\bibinfo {author} {\bibfnamefont {Engin}\ \bibnamefont
  {Torun}}, \bibinfo {author} {\bibfnamefont {Henrique P.~C.}\ \bibnamefont
  {Miranda}}, \bibinfo {author} {\bibfnamefont {Alejandro}\ \bibnamefont
  {Molina-S\'anchez}}, \ and\ \bibinfo {author} {\bibfnamefont {Ludger}\
  \bibnamefont {Wirtz}},\ }\bibfield  {title} {\enquote {\bibinfo {title}
  {Interlayer and intralayer excitons in ${\mathrm{mos}}_{2}/{\mathrm{ws}}_{2}$
  and ${\mathrm{mose}}_{2}/{\mathrm{wse}}_{2}$ heterobilayers},}\ }\href
  {\doibase 10.1103/PhysRevB.97.245427} {\bibfield  {journal} {\bibinfo
  {journal} {Phys. Rev. B}\ }\textbf {\bibinfo {volume} {97}},\ \bibinfo
  {pages} {245427} (\bibinfo {year} {2018})}\BibitemShut {NoStop}%
\bibitem [{\citenamefont {Chan}\ \emph {et~al.}(2023)\citenamefont {Chan},
  \citenamefont {Haber}, \citenamefont {Naik}, \citenamefont {Neaton},
  \citenamefont {Qiu}, \citenamefont {da~Jornada},\ and\ \citenamefont
  {Louie}}]{Chan2023}%
  \BibitemOpen
  \bibfield  {author} {\bibinfo {author} {\bibfnamefont {Yang-hao}\
  \bibnamefont {Chan}}, \bibinfo {author} {\bibfnamefont {Jonah~B.}\
  \bibnamefont {Haber}}, \bibinfo {author} {\bibfnamefont {Mit~H.}\
  \bibnamefont {Naik}}, \bibinfo {author} {\bibfnamefont {Jeffrey~B.}\
  \bibnamefont {Neaton}}, \bibinfo {author} {\bibfnamefont {Diana~Y.}\
  \bibnamefont {Qiu}}, \bibinfo {author} {\bibfnamefont {Felipe~H.}\
  \bibnamefont {da~Jornada}}, \ and\ \bibinfo {author} {\bibfnamefont
  {Steven~G.}\ \bibnamefont {Louie}},\ }\bibfield  {title} {\enquote {\bibinfo
  {title} {Exciton lifetime and optical line width profile via exciton-phonon
  interactions: Theory and first-principles calculations for monolayer mos2},}\
  }\href {\doibase 10.1021/acs.nanolett.3c00732} {\bibfield  {journal}
  {\bibinfo  {journal} {Nano Lett.}\ } (\bibinfo {year} {2023}),\
  10.1021/acs.nanolett.3c00732}\BibitemShut {NoStop}%
\bibitem [{\citenamefont {Toyozawa}(1958)}]{Toyozawa1958}%
  \BibitemOpen
  \bibfield  {author} {\bibinfo {author} {\bibfnamefont {Yutaka}\ \bibnamefont
  {Toyozawa}},\ }\bibfield  {title} {\enquote {\bibinfo {title} {{Theory of
  Line-Shapes of the Exciton Absorption Bands}},}\ }\href {\doibase
  10.1143/PTP.20.53} {\bibfield  {journal} {\bibinfo  {journal} {Progress of
  Theoretical Physics}\ }\textbf {\bibinfo {volume} {20}},\ \bibinfo {pages}
  {53--81} (\bibinfo {year} {1958})},\ \Eprint
  {http://arxiv.org/abs/https://academic.oup.com/ptp/article-pdf/20/1/53/5457877/20-1-53.pdf}
  {https://academic.oup.com/ptp/article-pdf/20/1/53/5457877/20-1-53.pdf}
  \BibitemShut {NoStop}%
\bibitem [{SM()}]{SM}%
  \BibitemOpen
  \href@noop {} {}\bibinfo {note} {See Supplemental Material at [URL will be
  inserted by publisher].}\BibitemShut {Stop}%
\bibitem [{\citenamefont {Marini}(2008)}]{Marini2008}%
  \BibitemOpen
  \bibfield  {author} {\bibinfo {author} {\bibfnamefont {Andrea}\ \bibnamefont
  {Marini}},\ }\bibfield  {title} {\enquote {\bibinfo {title} {Ab initio
  finite-temperature excitons},}\ }\href {\doibase
  10.1103/PhysRevLett.101.106405} {\bibfield  {journal} {\bibinfo  {journal}
  {Phys. Rev. Lett.}\ }\textbf {\bibinfo {volume} {101}},\ \bibinfo {pages}
  {106405} (\bibinfo {year} {2008})}\BibitemShut {NoStop}%
\bibitem [{\citenamefont {Molina-S\'anchez}\ \emph {et~al.}(2016)\citenamefont
  {Molina-S\'anchez}, \citenamefont {Palummo}, \citenamefont {Marini},\ and\
  \citenamefont {Wirtz}}]{Molina2016}%
  \BibitemOpen
  \bibfield  {author} {\bibinfo {author} {\bibfnamefont {Alejandro}\
  \bibnamefont {Molina-S\'anchez}}, \bibinfo {author} {\bibfnamefont
  {Maurizia}\ \bibnamefont {Palummo}}, \bibinfo {author} {\bibfnamefont
  {Andrea}\ \bibnamefont {Marini}}, \ and\ \bibinfo {author} {\bibfnamefont
  {Ludger}\ \bibnamefont {Wirtz}},\ }\bibfield  {title} {\enquote {\bibinfo
  {title} {Temperature-dependent excitonic effects in the optical properties of
  single-layer ${\mathrm{mos}}_{2}$},}\ }\href {\doibase
  10.1103/PhysRevB.93.155435} {\bibfield  {journal} {\bibinfo  {journal} {Phys.
  Rev. B}\ }\textbf {\bibinfo {volume} {93}},\ \bibinfo {pages} {155435}
  (\bibinfo {year} {2016})}\BibitemShut {NoStop}%
\end{thebibliography}%

\end{document}